\documentclass[12pt,dvipdfmx]{article}

\usepackage{amsmath,amssymb,amsfonts,graphics,graphicx,amscd,amsfonts,epsf,epsfig,color}
\usepackage{here}
\usepackage{mediabb}
\newcommand {\beq}{\begin{equation}}
\newcommand {\eeq}{\end{equation}} 
\newcommand {\beqa}{\begin{eqnarray}}
\newcommand {\eeqa}{\end{eqnarray}}

\newcommand {\tr}{{\rm Tr\,}}

\newcommand {\ee}{\mbox{e}}
\newcommand{\ps}{\psi}

\newcommand{\bbR}{{\mathbb R}}

\allowdisplaybreaks[4]

\makeatletter
\@addtoreset{equation}{section}

\makeatother
\date{}
\begin{document}

\begin{flushright} 
%\today\\
 
% WIS/06/09-MAY-DPP\\ 
YITP-16-20\\
UTHEP-682\\
KEK-TH-1891
\end{flushright} 

\vspace{0.2cm}

\begin{center}
  {\LARGE
  
Numerical tests of the gauge/gravity duality conjecture 
\\[0.2cm]
for D0-branes at finite temperature and finite $N$
  }
\end{center}
%\vspace{0.1cm}
\vspace{0.1cm}
\begin{center}

         Masanori H{\sc anada}$^{abc}$\footnote
          {
 E-mail address : hanada@yukawa.kyoto-u.ac.jp},  
         Yoshifumi H{\sc yakutake}$^{d}$\footnote
          {
 E-mail address : yoshifumi.hyakutake.phys@vc.ibaraki.ac.jp},  %hyaku : e-mail changed
          Goro I{\sc shiki}$^{e}$\footnote
          {
 E-mail address : ishiki@het.ph.tsukuba.ac.jp} 
and
         Jun N{\sc ishimura}$^{fg}$\footnote
          {
 E-mail address : jnishi@post.kek.jp}

%\vspace{0.5cm}
\vspace{0.2cm}

$^a${\it Stanford Institute for Theoretical Physics,
Stanford University, Stanford, CA 94305, USA}

$^b${\it Yukawa Institute for Theoretical Physics, Kyoto University,\\
Kitashirakawa Oiwakecho, Sakyo-ku, Kyoto 606-8502, Japan}

$^c${\it The Hakubi Center for Advanced Research, Kyoto University,\\
Yoshida Ushinomiyacho, Sakyo-ku, Kyoto 606-8501, Japan}

$^d${\it College of Science, Ibaraki University,
Bunkyo 2-1-1, Mito, Ibaraki 310-8512, Japan} 

$^e${\it Center for Integrated Research 
in Fundamental Science and Engineering (CiRfSE),\\
Faculty of Pure and Applied Sciences,
University of Tsukuba,\\
Tsukuba, Ibaraki 305-8571, Japan}

$^f${\it KEK Theory Center, 
High Energy Accelerator Research Organization,\\
1-1 Oho, Tsukuba, Ibaraki 305-0801, Japan} 

$^g${\it Graduate University for Advanced Studies (SOKENDAI),\\
1-1 Oho, Tsukuba, Ibaraki 305-0801, Japan} 

\end{center}
%\newpage

%\vspace{1.5cm}
\vspace{0.5cm}

\begin{center}
  {\bf abstract}
\end{center}

\noindent According to the gauge/gravity duality conjecture, 
the thermodynamics of gauge theory describing D-branes 
corresponds to that of black branes in superstring theory. 
We test this conjecture directly
in the case of D0-branes 
by applying Monte Carlo methods to
the corresponding gauge theory, which takes the form of 
the BFSS matrix quantum mechanics.
%% The obtained results are compared with
%% the corresponding quantity on the gravity side
%% including the $\alpha '$ and string loop corrections.
%
%and by comparing the results with the black 0-brane thermodynamics.
In particular, we take the continuum limit by extrapolating the
UV cutoff to infinity.
%$\infty$.
First we perform simulations at large $N$ so that
string loop corrections can be neglected on the gravity side.
%We have improved our previous analysis by taking the continuum limit.
Our results for the internal energy exhibit
the temperature dependence consistent with the prediction
%from the gravity side 
including the $\alpha '$ corrections.
Next we perform simulations at small $N$ but at lower temperature
so that the $\alpha '$ corrections can be neglected on the gravity side.
%The results obtained on the gauge theory side
Our results are consistent with the prediction
%from the gravity side 
including the leading string loop correction,
which suggests that the conjecture
%the gauge/gravity duality conjecture
holds even at finite $N$.
%In both parameter regions, we have taken the continuum limit
%
%up to the leading order of $1/N$.
%including $1/N$ corrections.

%holds beyond the large-$N$ limit.

%%%%%%%%%%%%%%%%%%%%%%%%%%%%%%%%%%%%%%%%%%%%%%%%%%%%%%%%%%%%%%%%%%%%%%
%%%%%%%%%%%%%%%%%%%%%%%%%%%%%%%%%%%%%%%%%%%%%%%%%%%%%%%%%%%%%%%%%%%%%%
%%%%%%%%%%%%%%%%%%%%%%%%%%%%%%%%%%%%%%%%%%%%%%%%%%%%%%%%%%%%%%%%%%%%%%
\newpage

%%%%%%%%%%%%%%%%%%%%%%%%%%%%%%%%%%%%%%%%%%%%%%%%%%%%%%%%%%%%%%%%%%%%%%
%%%%%%%%%%%%%%%%%%%%%%%%%%%%%%%%%%%%%%%%%%%%%%%%%%%%%%%%%%%%%%%%%%%%%%
%%%%%%%%%%%%%%%%%%%%%%%%%%%%%%%%%%%%%%%%%%%%%%%%%%%%%%%%%%%%%%%%%%%%%%
\section{Introduction}
%\hspace{0.51cm}
%%%%%%%%%%%%%%%%%%%%%%%%%%%%%%%%%%%%%%%%%%%%%%%%%%%%%%%%%%%%%%%%%%%%%%
%%%%%%%%%%%%%%%%%%%%%%%%%%%%%%%%%%%%%%%%%%%%%%%%%%%%%%%%%%%%%%%%%%%%%%
%%%%%%%%%%%%%%%%%%%%%%%%%%%%%%%%%%%%%%%%%%%%%%%%%%%%%%%%%%%%%%%%%%%%%%

One of the most important directions in theoretical physics is 
to clarify the quantum nature of gravity,
which is crucial in understanding the beginning of our Universe 
and the final state of a black hole.
%It is not just of academic interest; 
%the Higgs mass measured in LHC suggests that the standard model would be valid all the way up to the Planck scale, 
%and then one must understand quantum gravity in order to go beyond the standard model. 
%Also, the primordial gravitational wave which was claimed to be found recently, if true, 
%would provide us with direct access to quantum gravity. 
Superstring theory is considered the most promising candidate 
of a quantum gravity theory
due to its UV finiteness
%behavior of the theory is well controlled
in striking contrast to the conventional 
field theoretical approach to quantum gravity,
in which one faces with nonrenormalizable UV divergences.
So far, superstring theory is defined only perturbatively 
around simple backgrounds such as flat spacetime,
and it does not seem to be straightforwardly applicable to 
the studies of a strongly gravitating spacetime
such as the black hole geometry.
However, this difficulty has been partly surmounted
by the discovery of D-branes \cite{Polchinski:1995mt}. 
Some extremal %hyaku : typo
black holes were constructed by combining different kinds 
of D-branes, and the origin of their entropy was understood 
by counting the microstates of the D-branes \cite{Strominger:1996}.
Also there are several proposals for nonperturbative formulations
of superstring theory 
% in a background independent manner 
based on super Yang-Mills theory 
in low dimensions \cite{Banks:1996vh,Ishibashi:1996xs,Dijkgraaf:1997vv,Itoyama:1998et}.%hyaku : added Itoyama-Tokura
%from which some gravitational aspects of superstring theory 
%have been successfully extracted.

%% It is therefore expected that 
%% the quantum nature of gravity can be revealed 
%% by studying the Yang-Mills theory. 
In fact, it is conjectured that 
superstring theory on the anti-de Sitter background is 
dual to four-dimensional ${\cal N}=4$ 
U($N$) 
super Yang-Mills theory \cite{Maldacena:1997re},
which is realized on a stack of $N$ D3-branes.
%
%, which is called the AdS/CFT correspondence 
This duality has been generalized in various ways,
and it is commonly referred to as
the gauge/gravity duality conjecture.
The conjecture appears natural considering that D-branes
have two descriptions, one from a gravitational viewpoint,
and the other from a field theoretical viewpoint.
%% we can analyze 
%% both from the gravitational and field theoretical viewpoints.
If this conjecture is true even in the presence of 
quantum effects on the gravity side,
the quantum nature of gravity can be studied 
on a firm ground by investigating the dual gauge theory.
%super Yang-Mills theory. 

Among various gauge/gravity duality conjectures
that have been proposed so far, we are interested in the one 
that has been studied most intensively, which claims that
type II superstring theory in the near horizon limit of 
the black $p$-brane geometry is equivalent to 
a maximally supersymmetric 
%U($N$) 
Yang-Mills theory 
in $(p+1)$-dimensions \cite{Maldacena:1997re,Itzhaki:1998dd}. 
The super Yang-Mills theory is realized on $N$ D$p$-branes 
and it is characterized by the rank of the gauge group $N$ and 
the 't Hooft coupling $\lambda = g_{\text{YM}}^2 N$.
On the gravity side, $N$ is the number of black $p$-branes,
and the 't Hooft coupling is written 
as $\lambda = (2\pi)^{p-2} \ell_s^{p-3} g_s N$
in terms of the string coupling constant $g_s$ and the string length 
$\ell_s = \alpha'{}^{1/2}$.
The near horizon limit is taken by $\alpha' \to 0$
with $\lambda$ and the energy scale $U$ 
of the super Yang-Mills theory kept finite.
The above gauge/gravity duality has been tested 
in detail at $N\to\infty$ and $\lambda\to\infty$.
In this limit, superstring theory is well approximated 
by supergravity, which makes classical analyses applicable.
While the super Yang-Mills theory becomes strongly coupled
in this region, one can nevertheless extract
the information of BPS states, which are protected by supersymmetry,
and confirm the gauge/gravity duality in various ways.
%holds for $N\to\infty$ and $\lambda\to\infty$ region.

A natural question to ask then is 
whether the gauge/gravity duality conjecture is 
valid even for finite $\lambda$ or finite $N$,
or in a non-supersymmetric setup such as finite temperature.
%at finite temperature, which breaks supersymmetry.ic situation such as 
The analyses in these cases are quite difficult, however, 
because the supergravity approximation is no longer valid on the gravity side
and we have to include $\alpha'$ or $g_s$ corrections.
Furthermore, the lack of supersymmetry 
%admits
gives rise to 
all kinds of nonperturbative corrections to physical quantities
on the gauge theory side, which are too hard to handle analytically.

The main purpose of this paper is to test the gauge/gravity duality 
conjecture for finite $N$ and $\lambda$ 
at finite temperature.
%without supersymmetry.
While it is not possible to calculate
finite $\alpha'$ or $g_s$ corrections in superstring theory in general,
the corrections to some physical quantities can be extracted 
by taking account of higher
derivative corrections to supergravity perturbatively.
For instance, in the case of D0-branes,
the internal energy including the leading $g_s$ corrections 
has recently been evaluated analytically \cite{H2,H3}.
On the other hand, nonperturbative studies of the super Yang-Mills theory 
are possible by performing Monte Carlo simulation.
% of Yang-Mills theory plays an important role. 
%hyaku : replaced 'In particular' with 'Especially' 
In the case of D0-branes,
the super Yang-Mills theory takes the form of matrix quantum mechanics
for M theory \cite{Banks:1996vh,de Wit:1988ig}, 
which can be studied with reasonable amount of computational effort.
%Therefore we test the gauge/gravity duality conjecture for 
%the $p=0$ case by studying the quantum mechanics of $N$ D0-branes.
In fact, several groups have studied this model 
\cite{Kabat:1999hp,Kabat:2000zv,Kabat:2001ve,Iizuka:2001cw,%
Anagnostopoulos:2007fw,Hanada:2008gy,Hanada:2008ez,%
Hanada:2011fq,Hanada:2009ne,%
Hanada:2013rga,%
Catterall:2008yz,Catterall:2009xn,%
%Kadoh:2012bg,Kadoh:2014hsa
Kadoh:2015mka,Filev:2015hia}
and compared the obtained results with the dual gravity 
predictions. 
%\cite{Kabat:1999hp}-\cite{Kadoh:2014hsa}. 
%In all these papers, the comparison focused on the 
%large-$N$ limit, where
%quantum corrections in superstring theory are suppressed.
%hyaku : replaced 'In particular' with 'Among these works'
In particular, finite $\lambda$ corrections
%, however, were studied
were investigated by Monte Carlo simulation
in refs.~\cite{Anagnostopoulos:2007fw,Hanada:2008ez}
and more recently in refs.~\cite{Kadoh:2015mka,Filev:2015hia}, 
which raised some controversies.
In this paper, we first address this issue
based on new calculation, which improves our previous 
analysis \cite{Anagnostopoulos:2007fw,Hanada:2008ez}
by taking the continuum limit.

Then we investigate the $1/N$ corrections
by simulating the same system at small $N$.
This turns out to be much more difficult than
the studies at large $N$ because of the instability associated
with the flat directions in the potential.
%action.
The bound state of D0-branes is stable at large $N$, 
but it becomes only meta-stable at sufficiently low temperature
for small $N$.
We extract the internal energy of the meta-stable bound states
by introducing a cutoff on the extent of the D0-brane distribution,
which is chosen in such a way that the obtained result
does not depend on it within a certain region.
Our results obtained in this way turn out to be
consistent with the analytic result obtained on the gravity side
including the leading $g_s$ corrections.
%In fact, the specific heat of the analytic result becomes negative
%at low temperature, which 
%The meta-stability of the D0-brane bound state
%string loop effects.
This suggests that the gauge/gravity duality holds
even at finite $N$.
%including $1/N$ corrections.
In fact, the instability at finite $N$ can be understood
also on the gravity side.
Some of the results 
%at finite $N$ 
are reported briefly in our previous publication \cite{Hanada:2013rga}.

The rest of this paper is organized as follows.
In section \ref{sec:ov_gravity} we give an overview
of the black 0-brane thermodynamics,
%of higher derivative corrections in type II superstring theory,
and discuss how finite $\lambda$ and finite $N$ corrections
appear in the internal energy of the D0-branes.
In section \ref{sec:ov_QM} 
we explain how we study the D0-brane quantum mechanics 
by Monte Carlo simulation.
In section \ref{sec:numerical-tests}
we provide numerical tests of the 
gauge/gravity duality including finite $\lambda$ and finite $N$ corrections.
%% In section \ref{sec:large_N} 
%% we improve our previous analysis 
%% in the large-$N$ limit by taking the continuum limit carefully
%% and discuss the gauge/gravity duality including finite $\lambda$ corrections.
%% In section \ref{sec:finite_N} we present a new test of the
%% gauge/gravity duality including finite $N$ corrections.
Section \ref{sec:conclusion} is devoted to
a summary and discussions.
%future prospects.
%Conclusion and future directions are discussed in 

%%%%%%%%%%%%%%%%%%%%%%%%%%%%%%%%%%%%%%%%%%%%%%%%%%%%%%%%%%%%%%%%%%%%%%
%%%%%%%%%%%%%%%%%%%%%%%%%%%%%%%%%%%%%%%%%%%%%%%%%%%%%%%%%%%%%%%%%%%%%%
%%%%%%%%%%%%%%%%%%%%%%%%%%%%%%%%%%%%%%%%%%%%%%%%%%%%%%%%%%%%%%%%%%%%%%
\section{Brief review on the black 0-brane thermodynamics}
\label{sec:ov_gravity}
%\hspace{0.51cm}
%%%%%%%%%%%%%%%%%%%%%%%%%%%%%%%%%%%%%%%%%%%%%%%%%%%%%%%%%%%%%%%%%%%%%%
%%%%%%%%%%%%%%%%%%%%%%%%%%%%%%%%%%%%%%%%%%%%%%%%%%%%%%%%%%%%%%%%%%%%%%
%%%%%%%%%%%%%%%%%%%%%%%%%%%%%%%%%%%%%%%%%%%%%%%%%%%%%%%%%%%%%%%%%%%%%%

In this section we briefly review the thermodynamics of
the black 0-brane in type IIA superstring theory, which
appears in the gauge/gravity duality we are going to test.
In particular, we derive an expression for the quantity
that should be compared with the internal energy
of the dual gauge theory calculable by Monte Carlo methods.
Corresponding to finite $\lambda$ and finite $N$ corrections on the
gauge theory side, we need to consider
how the black 0-brane thermodynamics is affected by
the higher derivative corrections to the low-energy
effective action of type IIA superstring theory.

%%%%%%%%%%%%%%%%%%%%%%%%%%%%%%%%%%%%%%%%%%%%%%%%%%%%%%%%%%%%%%%%%%%%%%
%%%%%%%%%%%%%%%%%%%%%%%%%%%%%%%%%%%%%%%%%%%%%%%%%%%%%%%%%%%%%%%%%%%%%%
%%%%%%%%%%%%%%%%%%%%%%%%%%%%%%%%%%%%%%%%%%%%%%%%%%%%%%%%%%%%%%%%%%%%%%
\subsection{The effects of higher derivative corrections 
on the black 0-brane thermodynamics}
\label{sec:general-analysis}
%\hspace{0.51cm}
%%%%%%%%%%%%%%%%%%%%%%%%%%%%%%%%%%%%%%%%%%%%%%%%%%%%%%%%%%%%%%%%%%%%%%
%%%%%%%%%%%%%%%%%%%%%%%%%%%%%%%%%%%%%%%%%%%%%%%%%%%%%%%%%%%%%%%%%%%%%%
%%%%%%%%%%%%%%%%%%%%%%%%%%%%%%%%%%%%%%%%%%%%%%%%%%%%%%%%%%%%%%%%%%%%%%

In the low energy limit, the scattering amplitudes of strings 
in type II superstring theory 
can be reproduced correctly by
%the interactions in 
the type II supergravity action.
However, if we go beyond the low energy limit,
we have to take into account 
%the interactions of closed strings, which give rise to corrections due to 
the effects due to the finite length $\ell_s$
of strings ($\alpha'$ corrections) 
and 
%due to 
the effects of string loops ($g_s$ corrections).
%where we define $\alpha' = \ell_s^2$ and the string coupling constant $g_s$.
%
%and the deviations from supergravity becomes significant.
%hyaku : modified 2 sentences below and added Gross-Witten and Gross-Sloan
In general, these effects can be extracted by considering the scattering amplitudes
of strings associated with a Riemann surface with genus $n$
and expanding them with respect to external momenta \cite{Gross:1986iv,Gross:1986mw}.
The number of external momenta in the expansion raises the power of 
$\alpha'$, whereas the genus $n$ gives the power of $g_s^{2}$.
The corrections to the type II supergravity action due to these effects 
%hyaku : the effective action of superstring theory -> the type II supergravity action
%should include 
are represented by higher derivative terms,
and they are organized in the form of a double expansion
with respect to $\alpha'$ and $g_s$.
Below, we discuss some qualitative features of the higher derivative terms
in type IIA superstring theory.

%% In order to reproduce the S-matrix for closed strings, 
%% the effective action of type IIA superstring 
%% theory should include higher derivative terms. 
Treating the type IIA superstring theory
perturbatively with respect to 
two parameters $\alpha' = \ell_s^2$ and the dilaton coupling $g_s e^{\phi}$, 
one can write its effective action formally as
\begin{alignat}{3}
  S = \frac{1}{2\kappa_{10}^2} \int d^{10}x \sqrt{-g} e^{-2\phi}
  \Big( \mathcal{O}^{(0)} + 
\sum_{m,n} \ell_s^{2m} (g_s e^{\phi})^{2n} \mathcal{O}^{(m,n)} \Big) \ , 
  \label{eq:IIAact}
\end{alignat}
with an overall coefficient given by 
$2\kappa_{10}^2 = (2\pi)^7 \ell_s^8 g_s^2$.
Here $\mathcal{O}^{(0)}$ represents the terms with mass dimension 2 
that appear in the type IIA supergravity action,
and $\mathcal{O}^{(m,n)}$ represents higher derivative corrections 
with mass dimension $(2m+2)$.
All these terms are written in terms of the massless fields 
in the type IIA superstring theory
such as the graviton $g_{\mu\nu}$, the dilaton $\phi$
and the R-R 1-form potential $C_\mu$.
The structure of the higher derivative terms can be determined by
explicit calculations of scattering amplitudes,
which show that
%and it is known that
the $\ell_s^6$ and $\ell_s^6 g_s^2$ terms appear as
the leading corrections, respectively, at the tree level and at the one-loop level.
It is also known that these terms 
%are not modified by higher-loop effects.
do not appear from higher loops.
Thus we only have terms in eq.~(\ref{eq:IIAact})
with $m \geq 3$ for $n = 0,1$ and 
with $m > 3$ for $n \geq 2$ \cite{GRV}.
%with $m \geq n+3$ for $n \geq 2$ \cite{GRV}.

The equations of motion that one obtains
from the effective action (\ref{eq:IIAact})
can also be expanded in a power series as
%hyaku : replaced E with \mathcal{E}, since E is used for internal energy
\begin{alignat}{3}
  \mathcal{E} &= \mathcal{E}^{(0)} + 
\sum_{m,n} \ell_s^{2m} (g_s e^{\phi})^{2n} \mathcal{E}^{(m,n)} = 0 \ , 
\label{eq:eomIIA}
\end{alignat}
omitting the tensor indices for simplicity.
Here $\mathcal{E}^{(0)}$ represents the part obtained from the type IIA supergravity, 
and $\mathcal{E}^{(m,n)}$ %hyaku : E -> \mathcal{E}
%has mass dimension $(2m+2)$. 
represents the part obtained from the high derivative corrections.
%Here tensor indices are omitted for simplicity.
In order to solve the above equations of motion for the black 0-brane,
we make a general ansatz for $g_{\mu\nu}$, $\phi$ and $C_\mu$
respecting SO(9) rotational symmetry as \cite{H2}
\begin{alignat}{3}
  &ds^2 = - H_1^{-1} H_2^{\frac{1}{2}} F_1 dt^2 
  + H_2^{\frac{1}{2}} F_1^{-1} dr^2 
  + H_2^\frac{1}{2} r^2 d\Omega_8^2 \ , \notag
  \\[0.2cm]
  &e^\phi = H_2^\frac{3}{4}, \qquad
  C = \sqrt{1 + \alpha^7} (H_2 H_3)^{-\frac{1}{2}} dt \ , \label{eq:IIAsol}
  \\[0.1cm]
  &H_i(r) = 1 + \frac{r_-^7}{r^7} 
  + \sum_{m,n} \ell_s^{2m} g_s^{2n} H_i^{(m,n)}(r) \ , \notag
  \\
  &F_1(r) = 1 - \frac{(r_- \alpha)^7}{r^7} 
  + \sum_{m,n} \ell_s^{2m} g_s^{2n} F_1^{(m,n)}(r) \ , \notag
\end{alignat}
which involves four unknown functions $H_i(r)\,(i=1,2,3)$ and $F_1(r)$.
The leading behaviors of $H_i$ and $F_1$ are fixed
by using the solution of $\mathcal{E}^{(0)} = 0$ which are asymptotically flat, 
%hyaku : E -> \mathcal{E}
and they involve two parameters $r_-$ and $\alpha$.
As we will see below, $r_-$ and $\alpha$ are related to %hyaku : added 1 sentence
the mass and the charge of the black 0-brane. 
The subleading terms described by the functions $H_i^{(m,n)}$ and $F_1^{(m,n)}$
can be obtained by solving eq.~(\ref{eq:eomIIA}) order by order.
Note that 
%$H_i^{(3,0)}$ and $H_i^{(3,1)}$ 
the functions for $(m,n)=(3,0)$ and $(m,n)=(3,1)$ can be obtained independently 
of the other unknowns
since they are the leading corrections, respectively, 
at the tree level and at the one-loop level.
%On the other hand, $H_i^{(3m,0)}$, for example, is affected by 
%$H_i^{(3,0)}, \cdots, H_i^{(3(m-1),0)}$ iteratively. 

The event horizon $r_\text{H}$, 
which is defined by $F_1(r_\text{H})=0$, 
can be obtained perturbatively as 
$r_\text{H} = r_- \alpha + \cdots$.
%once the function $F_1(r)$ is 
Then the Hawking temperature $\tilde{T}$ can be obtained
by requiring the absence of
%so as to remove 
conical singularity in the Euclidean geometry at the event horizon
as\footnote{Here we reserve the variables such as $T$,
$E$ and $U_0$
without tildes for dimensionless quantities to be defined in (\ref{eq:dim-less}).}
\begin{alignat}{3}
  \tilde{T} &= 
\left. 
\frac{1}{4\pi} H_1^{-\frac{1}{2}} \frac{dF_1}{dr} 
\right|_{r_\text{H}}
  = \frac{7(r_- \alpha)^\frac{5}{2}}{4\pi r_-^\frac{7}{2} \sqrt{1+\alpha^7}}
  \left( 1 + \sum_{m,n} \ell_s^{2m} g_s^{2n} \tilde{T}^{(m,n)} \right) \ ,
\label{eq:IIAT}
\end{alignat}
where $\tilde{T}^{(m,n)}$ can be determined once the solution is obtained.
Since the metric (\ref{eq:IIAsol}) is asymptotically flat, 
the mass $\tilde{M}$ of the black 0-brane
can be evaluated by using the ADM mass formula and 
the charge $\tilde{Q}$ can be calculated by integrating 
the R-R flux\footnote{The integrands for the ADM mass and the R-R charge 
have corrections due to the higher derivative terms,
which vanish at $r=\infty$ \cite{H3}.}. 
They can be written formally as
\begin{alignat}{3}
  \tilde{M} &= \frac{V_{S^8}}{2\kappa_{10}^2} (r_- \alpha)^7
  \left( \frac{7 + 8 \alpha^7}{\alpha^7} + 
\sum_{m,n} \ell_s^{2m} g_s^{2n} \tilde{M}^{(m,n)} \right) \ , \label{eq:IIAMQ}
  \\
  \tilde{Q} &= \frac{V_{S^8}}{2\kappa_{10}^2} (r_- \alpha)^7 
  \left( \frac{7 \sqrt{1 + \alpha^7}}{\alpha^7} 
+ \sum_{m,n} \ell_s^{2m} g_s^{2n} \tilde{Q}^{(m,n)} \right) \ , 
\label{eq:IIAMQ2}
%\notag
\end{alignat}
where $V_{S^8} = \frac{2 \pi^{9/2}}{\Gamma(9/2)} = 
\frac{2(2\pi)^4}{7\cdot 15}$ is the volume of $S^8$.
%$\tilde{M}^{(m,n)}$ and $\tilde{Q}^{(m,n)}$ are expansion coefficients.
The internal energy of $N$ D0-branes
$\tilde{E} = \tilde{M} - \tilde{Q}$,
which is identified as the difference between the mass and the charge, 
can be obtained from (\ref{eq:IIAMQ}) and (\ref{eq:IIAMQ2}) as
\begin{alignat}{3}
  \tilde{E} &= \frac{V_{S^8}}{2\kappa_{10}^2} (r_- \alpha)^7
  \left( \frac{1 + 8 \sqrt{1 + \alpha^7}}{1 + \sqrt{1 + \alpha^7}} 
  + \sum_{m,n} \ell_s^{2m} g_s^{2n} \tilde{E}^{(m,n)} \right) \ , 
\label{eq:IIAE}
\end{alignat}
where $\tilde{E}^{(m,n)}$ has mass dimension $2m$.
%% Clearly the stringy and loop corrections affect the internal energy, 
%% and in some region
%% these effects become comparable to the classical value.
%% As explained in appendix \ref{app:gravity}, 
%% it is possible to evaluate $\tilde{E}^{(3,1)}$ explicitly.

%%%%%%%%%%%%%%%%%%%%%%%%%%%%%%%%%%%%%%%%%%%%%%%%%%%%%%%%%%%%%%%%%%%%%%
%%%%%%%%%%%%%%%%%%%%%%%%%%%%%%%%%%%%%%%%%%%%%%%%%%%%%%%%%%%%%%%%%%%%%%
%%%%%%%%%%%%%%%%%%%%%%%%%%%%%%%%%%%%%%%%%%%%%%%%%%%%%%%%%%%%%%%%%%%%%%
\subsection{Black 0-brane thermodynamics at large $N$}
\label{sec:large-N_gravity_side}
%\hspace{0.51cm}
%%%%%%%%%%%%%%%%%%%%%%%%%%%%%%%%%%%%%%%%%%%%%%%%%%%%%%%%%%%%%%%%%%%%%%
%%%%%%%%%%%%%%%%%%%%%%%%%%%%%%%%%%%%%%%%%%%%%%%%%%%%%%%%%%%%%%%%%%%%%%
%%%%%%%%%%%%%%%%%%%%%%%%%%%%%%%%%%%%%%%%%%%%%%%%%%%%%%%%%%%%%%%%%%%%%%

Let us first consider 
%the black 0-brane solution within 
the supergravity approximation,
which is valid
%This is valid 
when the curvature radius is large compared to $\ell_s$
and the effective coupling $g_s e^\phi$ is small.
Neglecting higher derivative corrections 
in eqs.~(\ref{eq:IIAT}) and (\ref{eq:IIAE}),
the temperature and the internal energy of the black 0-brane are 
obtained as
\begin{alignat}{3}
  \tilde{T} = \frac{7(r_- \alpha)^\frac{5}{2}}{4\pi r_-^\frac{7}{2} 
\sqrt{1+\alpha^7}} \ , \qquad
  \tilde{E} = \frac{V_{S^8}}{2\kappa_{10}^2} (r_- \alpha)^7
  \frac{1 + 8 \sqrt{1 + \alpha^7}}{1 + \sqrt{1 + \alpha^7}} \ , 
\label{eq:E}
\end{alignat}
where $r_-$ and $\alpha$ are the parameters of the classical black 0-brane.
%$2\kappa_{10}^2 = (2\pi)^7 \ell_s^8 g_s^2$ and
%$V_{S^8} = \frac{2 \pi^{9/2}}{\Gamma(9/2)} = \frac{2(2\pi)^4}{7\cdot 15}$ 
%is the volume of $S^8$.
%% The extremal limit corresponds to
%% $\alpha \rightarrow 0$,
%% which implies
%% $r_-^7 \rightarrow  (2\pi)^215\pi g_s N \ell_s^7$
%% due to $\tilde{Q}=N/(\ell_s g_s)$ and (\ref{eq:IIAMQ2}).
%hyaku : added the explanation for \tilde{Q}
The extremal limit corresponds to
$\alpha \rightarrow 0$
and $r_-^7 \rightarrow  (2\pi)^215\pi g_s N \ell_s^7$,
where the latter follows from the former 
using $\tilde{Q}=N/(\ell_s g_s)$ and (\ref{eq:IIAMQ2}).
In that limit,
the event horizon $r_\text{H} = r_- \alpha$, the temperature $\tilde{T}$
and the internal energy $\tilde{E}$ all vanish
as long as $r_-$ is kept finite.

%In order to compare with the results of the D0-brane quantum mechanics, 
The gauge/gravity duality holds in
%we need to take 
the near horizon limit \cite{Maldacena:1997re,Itzhaki:1998dd},
which is given in the present case by
\begin{alignat}{3}
  \ell_s \to 0 \quad \text{with} \quad 
\tilde{U}_0 \equiv \frac{r_\text{H}}{\ell_s^2} \quad \text{and} \quad
  \lambda = \frac{g_s N}{(2\pi)^2 \ell_s^3} \quad \text{fixed.}
\label{lambda-def-grav}
\end{alignat}
Note that the gravitational coupling $\kappa_{10}^2$ goes to zero
when $\ell_s \to 0$ with the 't Hooft coupling $\lambda = g_\text{YM}^2 N$ fixed.
This means that the gauge theory on the D0-branes decouples from the bulk gravity.
%In (\ref{lambda-def-grav}), we also require $\tilde{U}_0$ to be finite.
On the other hand, the parameter $\tilde{U}_0$ 
in (\ref{lambda-def-grav})
is proportional 
to the product of the string tension 
$1/2\pi \ell_s^2$ and the typical length $r_\text{H}$, 
which represents the gauge boson mass in the gauge theory.
Therefore, fixing $\tilde{U}_0$ in the limit corresponds
to keeping the energy scale of the dual gauge theory finite.
Let us also mention that the limit (\ref{lambda-def-grav})
can be rewritten
in terms of $\alpha$ and $r_-$,
%, the near horizon limit is reexpressed as
as
\begin{alignat}{3}
  \alpha \to 0 \ , \qquad 
\frac{r_-^2}{\alpha^5} \to (2\pi)^4 15\pi \lambda \tilde{U}_0^{-5}\ , \qquad
  \frac{2\kappa_{10}^2}{(r_-\alpha)^7} \to 
\frac{(2\pi)^{11}\lambda^2 \tilde{U}_0^{-7}}{N^2} \ .
\label{near-horizon-limit}
\end{alignat}
Since $\alpha \to 0$ and $r_- \to 0$,
the near horizon limit may be regarded as
a special case of the near extremal limit,
in which the temperature $\tilde{T}$
and the internal energy $\tilde{E}$ are kept finite.

In the near horizon limit, 
physical quantities can be 
expressed in terms of $\tilde{U}_0$ and $\lambda$.
Introducing a rescaled coordinate %$U = \frac{r}{\ell_s^2}$,
$U = r/ \ell_s^2$,
we can rewrite 
%the metric of the black 0-brane 
the solution (\ref{eq:IIAsol}) as \cite{Itzhaki:1998dd}
\begin{alignat}{3}
  &ds^2 = \ell_s^2 \big( - H^{-\frac{1}{2}} F dt^2 + 
   H^{\frac{1}{2}} F^{-1} dU^2
  + H^\frac{1}{2} U^2 d\Omega_8^2 \big) \ , \notag
  \\
  &e^{\phi} = \ell_s^{-3} H^\frac{3}{4} \ , 
  \quad C = \ell_s^4 H^{-1} dt \ , \label{eq:nearH}
  \\
  &H = \frac{(2\pi)^4 15\pi \lambda}{U^7} \ , \quad 
  F = 1 - \frac{\tilde{U}_0^7}{U^7} \ . \notag
\end{alignat}
%temperature $T = \tilde{T}/\lambda^\frac{1}{3}$ 
%and the energy scale $U_0 = \tilde{U}_0/\lambda^\frac{1}{3}$.
%We also define dimensionless internal energy $E = \tilde{E}/\lambda^\frac{1}{3}$, 
Taking the near horizon limit in eq.~(\ref{eq:E}), 
we obtain
\begin{alignat}{3}
  T & = a_1 U_0^\frac{5}{2} \ , \qquad
  a_1 = \frac{7}{16\pi^3 \sqrt{15\pi}} \ , 
\label{eq:T_sugra}
\\
  \frac{E}{N^2} & = \frac{18}{7^3} a_1^2 U_0^7 
  = \frac{18}{7^3} a_1^{-\frac{4}{5}} T^\frac{14}{5} 
\sim 7.41 \,T^{2.8} \ , \label{eq:E_sugra}
\end{alignat}
where we have defined
%Using the 
the dimensionless quantities
\begin{alignat}{3}
 T = \frac{\tilde{T}}{\lambda^\frac{1}{3}} \ , \quad \quad
 U_0 = \frac{\tilde{U}_0}{\lambda^\frac{1}{3}} \ , \quad \quad
 E = \frac{\tilde{E}}{\lambda^\frac{1}{3}} \ .
 \label{eq:dim-less}
\end{alignat}
Since the curvature radius $\rho$ and the effective coupling $g_s e^\phi$ 
around the event horizon are estimated as
\begin{alignat}{3}
  \frac{\ell_s}{\rho} \sim U_0^{\frac{3}{4}} 
%  \sim T^{\frac{3}{10}} 
\ , \qquad
  g_s e^\phi \sim \frac{1}{N} U_0^{-\frac{21}{4}} 
%\sim \frac{1}{N} T^{-\frac{21}{10}} 
\ , \label{eq:rho-gs}
\end{alignat}
the result (\ref{eq:E_sugra}) for the internal energy is valid 
when $U_0 \ll 1$ and $U_0^{-\frac{21}{4}} \ll N$,
which translates to $T \ll 1$ and $T^{-\frac{21}{10}} \ll N$
due to (\ref{eq:T_sugra}).

When $N$ is large but $T \sim O(1)$, 
%the effective coupling
%$g_s e^{\phi}$ becomes small and the quantum effects are negligible. 
$\ell_s / \rho$ is no longer small, and all the higher order terms
with $n=0$ remain in (\ref{eq:IIAT}) and (\ref{eq:IIAE}).
Therefore, eq.~(\ref{eq:E_sugra})
%$E/N^2$ depends only on $T$ and it can be expanded formally as 
should be replaced by
\begin{alignat}{3}
  \frac{E}{N^2} &= 7.41 \, T^{2.8} 
\left( 1 + \sum_{m \geq 3} c_{m,0} T^{\frac{3m}{5}} \right)
  = 7.41 \, T^{2.8} + a \, T^{4.6} + \tilde{a} \, T^{5.8} + \cdots 
\ , 
\label{eq:E_alpha}
\end{alignat}
where the second term
%$T^{4.6}$ 
comes from the leading $\alpha'$ correction with $(m,n)=(3,0)$
and the third term
comes from the next-leading $\alpha'$ correction with $(m,n)=(5,0)$.
The absence of $c_{4,0}$ follows from
some knowledge \cite{GRV} on the structure 
of the effective action (\ref{eq:IIAact}).
The explicit values of the non-zero coefficients $c_{m,0}$ 
are not known so far, however.

In fact, the internal energy of the black 0-brane 
is affected by the Hawking radiation, which 
has not been taken into account in (\ref{eq:IIAE}).
However, the energy loss through the Hawking radiation 
can be neglected in the near horizon limit,
as we show in appendix \ref{appendix:hawking}.
This is reasonable since 
the near horizon limit
implies the near extremal limit as well.
%as well.the near horizon limit corresponds to 

%% In general, the Wilson loop which involves the scalar field can 
%% be calculated from the gravity side 
%% by evaluating the area of the string worldsheet whose boundary is 
%% fixed by the Wilson loop \cite{Maldacena:1998im,Rey:1998ik}. 
%% For the D0-brane quantum mechanics, 
%% one can consider the Wilson loop winding on the temporal circle 
%% (the Polyakov loop), which is given by 
%% $W=\frac{1}{N}\text{Tr}{\rm P}e^{\int dt(iA+X)}$, 
%% where $X$ is one of the scalars $X_1,\cdots,X_9$. 
%% Its expectation value is evaluated as
%% \begin{eqnarray}
%%   \log\langle W\rangle \sim 1.89 \, T^{-3/5} + C \ , \label{eq:P_loop}
%% \end{eqnarray}
%% where $C$ is a constant.
%% Note that the above Polyakov loop is different from the ordinary one 
%% $P=\frac{1}{N}\text{Tr}{\rm P}e^{\int dt(iA)}$.
%% Nevertheless, the Polyakov loop $W$ is an order parameter for the $U(1)$ center symmetry along the 
%% temporal direction as usual.
%% Therefore, nonzero value of the loop means the system is in the deconfining phase 
%% at any nonzero temperature \cite{Witten:1998zw}. 

%%%%%%%%%%%%%%%%%%%%%%%%%%%%%%%%%%%%%%%%%%%%%%%%%%%%%%%%%%%%%%%%%%%%%%
%%%%%%%%%%%%%%%%%%%%%%%%%%%%%%%%%%%%%%%%%%%%%%%%%%%%%%%%%%%%%%%%%%%%%%
%%%%%%%%%%%%%%%%%%%%%%%%%%%%%%%%%%%%%%%%%%%%%%%%%%%%%%%%%%%%%%%%%%%%%%
\subsection{Black 0-brane thermodynamics at finite $N$}
%%%%%%%%%%%%%%%%%%%%%%%%%%%%%%%%%%%%%%%%%%%%%%%%%%%%%%%%%%%%%%%%%%%%%%
%%%%%%%%%%%%%%%%%%%%%%%%%%%%%%%%%%%%%%%%%%%%%%%%%%%%%%%%%%%%%%%%%%%%%%
%%%%%%%%%%%%%%%%%%%%%%%%%%%%%%%%%%%%%%%%%%%%%%%%%%%%%%%%%%%%%%%%%%%%%%

Let us move on to the case with finite $N$.
Since the effective coupling $g_s e^\phi$ 
given by (\ref{eq:rho-gs}) can no longer be neglected,
all the higher order terms
%with $(m,n)$ 
in eqs.~(\ref{eq:IIAT}) and (\ref{eq:IIAE}) remain.
Using (\ref{eq:rho-gs}),
each dimensionless term in (\ref{eq:IIAT}) and (\ref{eq:IIAE}) 
behaves as
\begin{alignat}{3}
  \ell_s^{2m} g_s^{2n} \tilde{E}^{(m,n)} 
   \sim \frac{1}{N^{2n}} U_0^{\frac{3m-21n}{2}} \ ,
\quad \quad
  \ell_s^{2m} g_s^{2n} \tilde{T}^{(m,n)} 
   \sim \frac{1}{N^{2n}} U_0^{\frac{3m-21n}{2}} \ .
\end{alignat}
Therefore, the internal energy of the black 0-brane 
in the near horizon limit is obtained as
\begin{alignat}{3}
  \frac{E}{N^2} = 7.41 \, T^{2.8} 
\left( 1 + \sum_{m,n} \frac{c_{m,n}}{N^{2n}} T^{\frac{3m-21n}{5}} \right) \ .
\label{general-formula}
\end{alignat}
The coefficients $c_{m,n}$ can be determined, in principle,
by solving the equations of motion (\ref{eq:eomIIA})
that can be derived from 
the explicit form of the effective action (\ref{eq:IIAact}).
%Similarly to the situation in the previous section,
Using some knowledge \cite{GRV}
on the structure of the effective action (\ref{eq:IIAact}),
we have $c_{m,1} \neq 0$ only for $m=3,6,\cdots$ and
$c_{m,2} \neq 0$ only for $m=5,6,\cdots$, etc..
%% Using some knowledge \cite{GRV}
%% on the structure of the effective action (\ref{eq:IIAact}),
%% we have $c_{m,0}\neq 0$ only for $m=3,5,6,\cdots$, 
%% $c_{m,1} \neq 0$ only for $m=3,6,\cdots$ and
%% $c_{m,2} \neq 0$ only for $m=5,6,\cdots$, etc..

In general, 
it is difficult to obtain the higher derivative corrections
in the effective action (\ref{eq:IIAact}).
There exists one exception, however, 
which is the leading correction at one loop
corresponding to $(m,n)=(3,1)$.
In this case, the correction can be obtained
\cite{H3} by uplifting the black 0-brane solution
to the M-wave solution in eleven dimensions, which
is purely geometrical.
Since higher curvature corrections in eleven dimensions are 
well-known \cite{Ts,BB,PVW,HO,H}, it is possible
to derive the equations of motion and solve them.
Using this result,
the internal energy including the leading $1/N^2$ correction 
can be obtained explicitly as \cite{H3} 
\begin{alignat}{3}
  \frac{E}{N^2} &= 
7.41 \, T^\frac{14}{5} - \frac{5.77}{N^2} \, T^{\frac{2}{5}} \ .
\label{eq:finite_N}
\end{alignat}
(See appendix \ref{appendix:one-loop} for a review on the derivation.)
%This is consistent with the result derived in ref.~\cite{H2}
%using the first law of the black hole thermodynamics.

Thus we find that the internal energy can be expanded 
with respect to $T$ and $1/N$ as
\begin{alignat}{3}
  \frac{E}{N^2} &= (7.41 \, T^{2.8} + a \, T^{4.6}+
  \tilde{a} \, T^{5.8}+\cdots) +
  (- 5.77 \, T^{0.4} + b \, T^{2.2}+\cdots)\frac{1}{N^2} + 
O\left(\frac{1}{N^4}\right) \ , 
\label{eq:IIAene}
\end{alignat}
where $a \,T^{4.6}$ and $\tilde{a} \,T^{5.8}$
correspond to the $\ell_s^6$ terms and the $\ell_s^{10}$ terms,
respectively, at the tree level, 
while $b \, T^{2.2}$ comes from the $\ell_s^{12} g_s^2$ terms 
at the one-loop level.
%% If we assume that all the coefficients 
%% in eq.~(\ref{eq:IIAene}) are $\mathcal{O}(1)$,
%% eq.~(\ref{eq:finite_N}) is reliable when 
%% $\frac{1}{\sqrt{0.1}T^{1.5}} \leq N \leq \frac{\sqrt{0.1}}{T^{2.1}}$ \cite{H2}.
%% This corresponds to 
%% $T^{0.3} \leq \sqrt{0.1} $ and $1000 \leq N$.
%% Therefore when we test the gauge/gravity duality conjecture around $N=5$, 
%% we have to take other terms in eq.~(\ref{eq:IIAene}) into account.

%%%%%%%%%%%%%%%%%%%%%%%%%%%%%%%%%%%%%%%%%%%%%%%%%%%%%%%%%%%%%%%%%%%%%%
%%%%%%%%%%%%%%%%%%%%%%%%%%%%%%%%%%%%%%%%%%%%%%%%%%%%%%%%%%%%%%%%%%%%%%
%%%%%%%%%%%%%%%%%%%%%%%%%%%%%%%%%%%%%%%%%%%%%%%%%%%%%%%%%%%%%%%%%%%%%%
%\section{Numerical method for D0-brane quantum mechanics} 
\section{D0-brane quantum mechanics} 
\label{sec:ov_QM}
%\hspace{0.51cm}
%%%%%%%%%%%%%%%%%%%%%%%%%%%%%%%%%%%%%%%%%%%%%%%%%%%%%%%%%%%%%%%%%%%%%%
%%%%%%%%%%%%%%%%%%%%%%%%%%%%%%%%%%%%%%%%%%%%%%%%%%%%%%%%%%%%%%%%%%%%%%
%%%%%%%%%%%%%%%%%%%%%%%%%%%%%%%%%%%%%%%%%%%%%%%%%%%%%%%%%%%%%%%%%%%%%%

According to the gauge/gravity duality conjecture,
the thermodynamics of the black 0-brane corresponds to that 
of the gauge theory describing the D0-brane,
%In order to test the conjecture, we therefore need to study
%the gauge theory.
which takes the form of the BFSS matrix quantum mechanics.
% in the D0-brane case.
In order to investigate the thermodynamics,
we use the Euclidean time and 
compactify it with the periodicity $\beta = 1/\tilde{T}$.
Then the action of D0-brane quantum mechanics 
at finite temperature $\tilde{T}$ is given by 
\begin{eqnarray}
  S = \frac{1}{g_{\text{YM}}^2} \int_0^{\beta} dt \, \text{Tr} \bigg\{ 
  \frac{1}{2} (D_t X_i)^2 - \frac{1}{4} [X_i , X_j]^2 
+ \frac{1}{2} \psi_\alpha D_t \psi_\alpha
  - \frac{1}{2} \psi_\alpha \gamma_i^{\alpha\beta} [X_i , \psi_\beta ] \bigg\} \ ,
\label{YMaction}
\end{eqnarray}
which can be obtained formally by dimensionally reducing 
the action of $(9+1)$d ${\cal N}=1$ U($N$) super Yang-Mills theory 
to $(0+1)$ dimension. 
We have introduced
$X_i(t)$  $(i=1,\cdots,9)$ and $\psi_\alpha(t)$
$(\alpha=1,\cdots , 16)$, which 
are bosonic and fermionic $N\times N$ Hermitian matrices, 
respectively, and the covariant derivative
$D_t  = \partial_t  - i \, [A(t), \ \cdot \ ]$
is defined using the ${\rm U}(N)$ gauge field $A(t)$. 
The bosonic variables obey periodic boundary conditions
$X_i(t+\beta)=X_i(t)$, $A(t+\beta)=A(t)$, whereas
the fermionic variables obey anti-periodic boundary conditions
$\psi_\alpha(t+\beta)=-\psi_\alpha(t)$.
The $16\times 16$ matrices $\gamma_i$ in (\ref{YMaction}) act on spinor indices and
satisfy the Euclidean Clifford algebra $\{ \gamma_i,\gamma_j \}= 2\delta_{ij}$.
%% where 
%% $X_i(t)$  $(i=1,\cdots,9)$ and $\psi_\alpha(t)$
%% $(\alpha=1,\cdots , 16)$ are bosonic and fermionic $N\times N$ hermitian matrices, 
%% respectively, and 
%% $D_t  = \partial_t  - i \, [A(t), \ \cdot \ ]$ is the covariant derivative with the $U(N)$ gauge field $A(t)$. 
%% The $16\times 16$ matrices $\gamma_i$ in (\ref{YMaction}) act on spinor indices and
%% satisfy the Euclidean Clifford algebra $\{ \gamma_i,\gamma_j \}= 2\delta_{ij}$.
%% This action (\ref{YMaction})
%% can be obtained formally by dimensionally reducing 
%% the action of $(9+1)$d ${\cal N}=1$ super Yang-Mills theory 
%% to $(0+1)$ dimension. 

The 't Hooft coupling $\lambda =g_{\text{YM}} ^2 N$
corresponds to $\lambda$ defined in (\ref{lambda-def-grav})
on the dual gravity side.
Since the coupling constant $g_{\text{YM}}^2$ 
in the action (\ref{YMaction}) has mass dimension 3,
all dimensionful quantities can be measured in units of $\lambda$
as we did in (\ref{eq:dim-less}).
Note that the expansion (\ref{eq:IIAene}) is valid when
$T \ll 1$ and $T^{-\frac{21}{10}} \ll N$.
%The second inequality implies that $N$ should be large with fixed $\lambda$,
%and the 
In particular, the
first inequality implies that $\lambda$ should be large for fixed
temperature $\tilde{T}$.
This implies that we need to study the strong coupling dynamics of
the D0-brane quantum mechanics in order to test the gauge/gravity duality.
For that purpose, we apply Monte Carlo methods analogous to the ones 
used in lattice QCD.

%\subsection{test at large $N$ --- $\alpha '$ corrections}
\subsection{Putting the theory on a computer}
\label{sec:putting-on-a-computer}

In order to apply Monte Carlo methods, we have to
put the D0-brane quantum mechanics (\ref{YMaction}) on a computer.
It is not possible to do it, however, respecting
all the maximal supersymmetry generated by 16 supercharges,
which the theory has at zero temperature.
For instance, if one discretizes the time direction,
one cannot maintain all the supersymmetry,
since successive supersymmetry transformations induce
a translation in time direction, which is broken by the discretization.
The lack of exact symmetry in quantum field theories 
typically necessitates fine tuning in taking the continuum limit
due to UV divergences.
This does not occur, however,
%it does not matter 
in the present case since the system is just 
a quantum mechanics, which is UV finite.
Here, instead of discretizing the time direction,
we expand the functions of $t$ in Fourier modes and
introduce a mode cutoff $\Lambda$ 
after fixing the gauge symmetry \cite{Hanada:2007ti}.
Since the higher Fourier modes omitted in our calculations
are suppressed by the kinetic term, the approach
to $\Lambda = \infty$ is expected to be fast.

We fix the gauge symmetry by the static diagonal gauge
\begin{eqnarray}
  A(t) = \frac{1}{\beta} \, {\rm diag} (\alpha_1 , \cdots , \alpha_N) \ ,
\end{eqnarray} 
where $\alpha_a$ are chosen to satisfy the constraint\footnote{In 
actual simulation, we replace the constraint by
$\max_a (\alpha_a) - \min_a (\alpha_a) \le 2\pi$.
This is practically equivalent to (\ref{pi-pi-constraint})
unless $\alpha_a$ is distributed in the whole region (\ref{pi-pi-constraint}),
which occurs only at very low temperature.}
\begin{eqnarray}
  - \pi < \alpha_a \le \pi \ .
\label{pi-pi-constraint}
\end{eqnarray} 
This constraint is needed to fix the symmetry under
large gauge transformations\footnote{The 
gauge transformation acts on the gauge field as 
$A(t)\to \Omega(t)^{-1} A(t)\Omega(t) + i \Omega(t)^{-1}\partial_t\Omega(t)$, 
where $\Omega(t)$ is an $N\times N$ unitary matrix 
which satisfies the periodic boundary condition 
$\Omega(t+\beta)=\Omega(t)$.   
After taking the static diagonal gauge, 
one still has to fix the residual symmetry under large gauge transformations,
which correspond to
$\Omega(t)={\rm diag}(e^{2\pi in_1 t/\beta},\cdots,e^{2\pi in_N t/\beta})$
with $n_1,\cdots,n_N$ being integers.}. 
The Faddeev-Popov term associated with this gauge fixing condition is given by
\begin{eqnarray}
  S_{\rm FP} = 
- \sum_{a<b} 2 \ln \left| \sin \frac{\alpha_a - \alpha_b}{2} \right| \ ,
\end{eqnarray}
which we add to the action (\ref{YMaction}).
The integration measure for $\alpha_a$ is taken to be uniform. 

Once we fix the gauge symmetry,
we can introduce the momentum cutoff $\Lambda$ 
for the Fourier modes of $X_i(t)$ and $\psi_\alpha(t)$. 
Since the bosonic matrices $X_i(t)$ obey periodic boundary conditions,
%$X_i(t+\beta)=X_i(t)$, 
they are expanded as 
\begin{eqnarray}
  X_i ^{ab} (t) = \sum_{n=-\Lambda}^{\Lambda} 
%\tilde{X}_{i n}^{ab} \, e^{2\pi i n t/\beta} \ ,  
\tilde{X}_{i n}^{ab} \, e^{i n \omega t} \ ,  
\end{eqnarray} 
where $\omega = 2 \pi / \beta$,
and $n$ runs over integers\footnote{Note that 
a large gauge transformation shifts the momentum 
of the mode $\tilde{X}_{i n}^{ab}$ as $n \to n-n_a+n_b$. 
Therefore, one needs to fix the symmetry under large gauge transformations
in order for the momentum cutoff to make sense.}. 
On the other hand, the fermionic matrices $\psi_\alpha(t)$, 
which obey anti-periodic boundary conditions, 
are expanded as
\begin{eqnarray}
  \psi_\alpha ^{ab} (t) = 
\sum_{r=-(\Lambda-1/2)}^{\Lambda-1/2} \tilde{\psi}_{\alpha r}^{ab} 
\, e^{ i  r \omega t} \ ,  
\end{eqnarray}
where $r$ runs over half-integers. 
%% In the remaining part of this section 
%% we consider the anti-periodic boundary condition. 
%% (A switch to the periodic boundary condition is straightforward.) 
By using a shorthand notation
\begin{eqnarray}
  \Bigl(f^{(1)}  \cdots  f^{(p)}\Bigr)_n \equiv 
\sum_{k_1 + \cdots + k_{p}=n} f^{(1)}_{k_1} \cdots f^{(p)}_{k_p} \ ,
\end{eqnarray}
the action \eqref{YMaction} can be expressed as $S=S_{\rm b}+S_{\rm f}$, where
\begin{alignat}{3}
  S_{\rm b} &= N \beta \Bigg[ \frac{1}{2} \sum_{n=-\Lambda}^{\Lambda} 
  \left( n \omega - \frac{\alpha_a - \alpha_b}{\beta} \right)^2 
\tilde{X}_{i,-n}^{ba} \tilde{X}_{i n}^{ab}
  - \frac{1}{4} \text{Tr} 
\Bigl( [ \tilde{X}_{i} , \tilde{X}_{j}]^2  \Bigr)_0 \Bigg] , 
%\nonumber 
  \label{bfss_action_cutoff_b}
  \\
  S_{\rm f} &= 
  \frac{1}{2} N \beta \Biggl[ \sum_{r=-(\Lambda-1/2)}^{\Lambda-1/2} 
 i \left(  r \omega - \frac{\alpha_a - \alpha_b}{\beta} \right)
  \tilde{\psi}_{\alpha , -r}^{ba} \tilde{\psi}_{\alpha r}^{ab} - 
   (\gamma_i)_{\alpha\beta}
  \text{Tr} \Bigl( \tilde{\psi}_{\alpha} 
  [ \tilde{X}_{i},\tilde{\psi}_{\beta}] \Bigr)_0  \Biggr] \ .
  \label{bfss_action_cutoff_f}
\end{alignat}  
This action is written in terms of 
a finite number of variables $\alpha_a$, $\tilde{X}_{i n}^{ab}$ 
and $\tilde{\psi}_{\alpha r}^{ab}$, and hence it can be dealt with on a computer. 
The continuum limit is realized by sending the cutoff $\Lambda$ to infinity. 
%Furthermore it is possible to reduce fermionic degrees of freedom.

The fermionic degrees of freedom are treated in the following way.
Note that the fermionic action $S_{\rm f}$ can be written as 
$S_{\rm f} = \frac{1}{2} 
{\cal M}_{A \alpha r , B \beta s} 
\tilde{\psi}_{\alpha r}^A \tilde{\psi}_{\beta s}^B$,
where we have expanded $\tilde{\psi}_{\alpha r}$ 
in terms of the generators $t^A$ of ${\rm U}(N)$ as
$\tilde{\psi}_{\alpha r}
= \sum_{A=1}^{N^2} \tilde{\psi}_{\alpha r}^A t^A$. 
By integrating out the fermionic variables, 
the partition function can be written as
\begin{eqnarray}
  Z = \int dX \, d\alpha \, d\psi \, e^{-S_{\rm b}-S_{\rm f}} =
  \int dX \, d\alpha \, {\rm Pf}{\cal M} \, e^{-S_{\rm b}} \ , 
\label{eq:Z}
\end{eqnarray}
where ${\rm Pf}{\cal M}$ represents the Pfaffian of ${\cal M}$, 
which is complex in general and is denoted as
${\rm Pf}{\cal M} =  |{\rm Pf}{\cal M}| \, e^{i\Gamma}$.
%% \footnote{The Pfaffian, rather than the determinant, 
%% appears because $\psi$ represents a Majorana fermion.}. 
Since Monte Carlo simulation is applicable only 
when the path integral has a positive semi-definite integrand, 
%we replace the Pfaffian by its absolute value,
we omit the phase factor $e^{i\Gamma}$
and define the expectation value of ${\cal O}(X,\alpha)$ 
for the phase-quenched model as
\begin{eqnarray}
  \Big\langle {\cal O}(X,\alpha) \Big\rangle_{\rm phase-quenched} \equiv
  \frac{\int dX \, d\alpha \, 
   {\cal O}(X,\alpha)
   |{\rm Pf}{\cal M}| \, e^{-S_{\rm b}} }{
  \int dX \, d\alpha \, |{\rm Pf}{\cal M}| \, e^{-S_{\rm b}}} \ . 
  \label{phase_quench}
\end{eqnarray} 
Then, the expectation value with respect to the original theory
(\ref{eq:Z}) is given by
\begin{eqnarray}
%  \Big\langle {\cal O}(X,\alpha) \Big\rangle_{\text{original}} =
  \Big\langle {\cal O}(X,\alpha) \Big\rangle = 
  \frac{\Big\langle {\cal O}(X,\alpha) \, e^{i\Gamma}  
\Big\rangle_{\rm phase-quenched}}{
  \Big\langle e^{i\Gamma} \Big\rangle_{\rm phase-quenched}} \ . 
  \label{phase_reweight}
\end{eqnarray} 
When $e^{i\Gamma}$ fluctuates rapidly
in 
%Monte Carlo simulation of (\ref{phase_quench}),
the simulation of the phase-quenched model,
it is difficult to evaluate (\ref{phase_reweight}) 
since both the denominator and the numerator become very small,
and the number of configurations needed to obtain the ratio
with sufficient accuracy becomes huge.
This technical problem is called the sign problem.

In the present system, however, it has been reported that
the fluctuation of $e^{i\Gamma}$ is strongly suppressed
at both high temperature and low temperature, 
and that it can be neglected throughout the whole temperature
region \cite{Catterall:2009xn,Filev:2015hia}.
This can be understood as follows.
At high temperature, the high temperature expansion 
becomes applicable \cite{Kawahara:2007ib},
which implies that the dynamics of $X$ and $\alpha$ in the Pfaffian
becomes perturbative.
Therefore, the fluctuation of $e^{i\Gamma}$ becomes less
important at high temperature.
%% At low temperature, 
At low temperature, on the other hand,
the dynamics is dominated by the low momentum modes,
for which the kinetic term (the first term) in (\ref{bfss_action_cutoff_f})
can be neglected.
If we omit the kinetic term, we can show that
the Pfaffian becomes real.
Thus, the effects of $e^{i\Gamma}$ can be neglected
also at low temperature, which is supported
by some numerical evidence \cite{Hanada:2011fq}.
%Thus, the fluctuation of $e^{i\Gamma}$ is expected to become 
%small at both high temperature and low temperature, and therefore,
In the present work, we simply omit the phase factor $e^{i\Gamma}$,
and use (\ref{phase_quench}) to calculate the VEV of observables.
%See appendix \ref{sec:simulation-detail} for some details of our simulation.
See appendix B of ref.~\cite{Hanada:2011fq}
for the details of our algorithm for Monte Carlo simulation.

Let us also comment on how we treat the zero modes.
Note that the constant modes $X_i (t) = x_i {\bf 1}_N$ ($x_i \in \bbR$)
of the trace part does not appear in the action (\ref{YMaction}).
These modes should be omitted in the path integral (\ref{eq:Z}).
%In what follows, we assume that the zero modes are fixed by the constraint
We can extract these modes from a general configuration by
\begin{align}
x_i =   \frac{1}{N\beta}  \int_0 ^\beta  dt \, {\rm Tr} \big(X_i(t) \big) \ .
\label{trX-constraint}
\end{align}
In what follows, we assume that these zero modes are fixed
by the constraint $x_i = 0$ for $i=1, \cdots , 9$.
In Monte Carlo simulation, even if we start from a configuration
satisfying the constraint, $x_i$ can become nonzero 
as the simulation proceeds due to accumulation of round-off errors.
%This is harmless as far as 
We avoid this by making a projection
$X_i (t) \mapsto X' _i (t) = X_i (t) - x_i {\bf 1}_N$.
%before we calculate quantities such as (\ref{R2-def})

%(\ref{trX-constraint})

%\subsection{test at large $N$ --- $\alpha '$ corrections}
\subsection{Calculation of the internal energy}
\label{sec:large-N}

The internal energy 
$E=-\frac{\partial }{\partial\beta} \ln Z$ 
of the D0-brane quantum mechanics can be calculated
using the formula
\begin{eqnarray}
  E = -3T \left(\langle 
S_\text{b}\rangle-\frac{9}{2}\Big\{(2\Lambda+1)N^2-1 \Big\}\right) \ ,
\label{energy-formula}
\end{eqnarray}
which can be obtained by adapting
the one used in the lattice formulation \cite{Catterall:2007fp}
to the present momentum cutoff formulation.
(See appendix \ref{sec:EO} for the derivation.)
%in the present momentum cutoff formulation.

A peculiar aspect of the D0-brane quantum mechanics
is that the action (\ref{YMaction}) has flat directions $[X_i,X_j]=0$.
These are lifted by quantum corrections
in the case of the bosonic model, in which fermionic degrees of freedom 
are omitted \cite{Kawahara:2007fn}.
However, in the supersymmetric model,
the flat directions are not lifted 
by quantum corrections due to supersymmetry.
% (at zero temperature).
As a result, the supersymmetric model
contains scattering states, 
which form the continuous branch of the energy spectrum,
in addition to the normalizable energy eigenstates,
which form the discrete branch of the 
spectrum.\footnote{In ref.~\cite{Smilga:2008bt},
the discrete branch of the spectrum is
shown to have a new energy scale proportional to $N^{-5/9}$
based on the effective Hamiltonian 
for the relevant ${\cal O}(N)$ degrees of freedom 
in the flat directions.
Based on this observation,
the particular power ``2.8''
of the leading behavior in (\ref{eq:E_alpha})
%$E/N^2 \sim 7.41 T^{14/5}$ 
has been understood theoretically
on the gauge theory side.
See also ref.~\cite{Hanada:2010jr} for related work on
supersymmetric models with 4 and 8 supercharges.}
Therefore, the path integral (\ref{eq:Z}) is 
ill-defined as it stands,
and we need to consider how to make sense out of it.
%in the context of the gauge/gravity duality.

%% First, in the high temperature limit, the fermions decouple and
%% the path integral (\ref{eq:Z}) is well-defined.
%% At high but finite temparature, the flat direction appears
%% only when the bosonic matrices $X_i$ take values larger than $O(T)$
%% due to the breaking of supersymmetry by 
%% the boundary conditions \cite{Anagnostopoulos:2007fw}.
%% Therefore, as far as $T$ is large enough, 
%% the path integral (\ref{eq:Z}) can be made well-defined.

In the large-$N$ limit, 
the path integral (\ref{eq:Z}) becomes well-defined
analogously to the well-known example 
of the $\Phi^3$ model \cite{Brezin:1977sv}.
In this case, 
%the normalizable states 
%actually decouple from the scattering states,
%and 
we may naturally consider that
the well-defined path integral (\ref{eq:Z})
actually represents the thermodynamics 
of the normalizable states only.
%in a standard manner.

The situation becomes subtle at finite $N$.
Suppose we prepare an initial state with $X_i \sim 0$
having sufficiently low energy
and let it evolve in time quantum mechanically.
%Below we describe what is deduced from the results
%of Monte Carlo simulation.
It is expected 
from the Monte Carlo simulation discussed in section \ref{sec:finite_N}
that the size ${\cal O} = \frac{1}{N}{\rm Tr} (X_i)^2$ of the state
fluctuates for a while
around some finite value depending on the initial state,
and eventually starts to diverge.
%Namely there seem to exist meta-stable bound states which have finite $R^2$.
%%
%% We consider the thermodynamics of these meta-stable states
%% which have finite size.
%%
%,which have finite ${\cal O}$.
%At finite $N$, the 
These meta-stable states 
are linear combinations
of normalizable states and scattering states.
%with the latter having much smaller amplitudes than the former.
%
%have certain components scattering states
%are not energy eigenstates of the original quantum mechanical system.
However, if 
%the amplitudes of the scattering states are suppressed and 
%the meta-stable states 
they are long-lived,
%and have a typical size for a given energy,
we can still think of their thermodynamics 
by introducing a cutoff ${\cal O} \le R^2_{\rm cut}$,
where $R^2_{\rm cut}$ should be chosen to be the 
typical size of the meta-stable states 
%with that energy.
for a given energy.
This can be achieved in the path integral formalism by
replacing the partition function (\ref{eq:Z}) by
\begin{eqnarray}
  Z = \int dX \, d\alpha \, d\psi \, e^{-S_{\rm b}-S_{\rm f}}
\theta(R^2_{\rm cut} - R^2_{\rm max})
=  \int dX \, d\alpha \, {\rm Pf}{\cal M} \, e^{-S_{\rm b}} 
\theta(R^2_{\rm cut} - R^2_{\rm max})
\ , 
\label{eq:Z-cutoff}
\end{eqnarray}
where $\theta(x)$ is the Heaviside step function and
we have defined
\begin{align}
  R^2_{\rm max} \equiv 
  \frac{1}{N}
  \max_{0 \le t \le \beta} \sum_{i=1}^9{\rm Tr} \big(X_i(t)^2 \big) \ .
\label{R2max-def}
\end{align}
It is expected that
the internal energy of the cutoff system (\ref{eq:Z-cutoff})
% obtained at some $T$ and $N$
becomes independent of $R^2_{\rm cut}$ within some region,
and the internal energy obtained in that region 
can be interpreted
as the average internal energy of the meta-stable states
for the 
%corresponding 
$T$ and $N$ corresponding to
the partition function (\ref{eq:Z-cutoff}).
%the lowest value in the region to represent
%the typical size of the meta-stable at that $T$ and $N$.
%%
%% However, we can still think of the thermodynamics 
%% if the internal energy of the cutoff system obtained 
%% at some $T$ and $N$
%% becomes independent of $R^2_{\rm cut}$ within some region.
The typical size of the meta-stable states
can be identified with the lower end of the 
region of $R^2_{\rm cut}$ within which the internal energy is constant.
%becomes independent.
%% Note that the size of the meta-stable depends on $T$,
%% ad therefore the internal energy defined above
%% is not necessarily a monotonously increasing function of $T$
%% unlike the situation in the large-$N$ limit.
Note that 
%the internal energy defined in this way
the partition function (\ref{eq:Z-cutoff})
does not represent the thermodynamics of
the normalizable states only unlike the case with $N=\infty$.
Therefore, there is no guarantee that 
%the internal energy becomes a monotonously increasing function of $T$.
the specific heat $C = \frac{d E}{dT}$ corresponding to
the internal energy $E$ 
defined in this way becomes positive.
%unlike the situation in the large-$N$ limit.

Since the $N$ eigenvalues of $X_i$ represent 
the position of the D0-branes,
the meta-stable states can be interpreted as
the bound states of D0-branes,
which form a black hole.
Therefore, it is expected that
the internal energy defined above corresponds to 
the internal energy (\ref{eq:IIAene}) of the black hole.
%On the gravity side, the black 
Indeed the black hole is stable in the large-$N$ limit,
but it becomes meta-stable at finite $N$ due to quantum effects
corresponding to $1/N^2$ corrections.
%This is reflected in the fact that
This can be seen in (\ref{eq:finite_N}), for instance, 
where the leading $1/N^2$ correction
% in (\ref{eq:finite_N})
makes the specific heat negative at sufficiently low $T$.
This instability can be understood physically
as caused by the repulsive force acting on
a test particle near the event horizon
due to the quantum gravity effects 
at small distances \cite{H2}.

%Note that the leading $1/N^2$ correction in
%(\ref{eq:finite_N}) is negative, and it 

%% It is consistent with the stringy picture.
%% Because the scalars correspond to the positions of the D0-branes, 
%% the meta-stable state represents the bound state of the D0-branes, 
%% which is regarded as the black 0-brane with some energy in the gravity picture. 

%% In Monte Carlo simulation, we prepare the initial configuration
%% of $X_i(t)$ by giving each element a small Gaussian random number.
%% As long as $N \gg N_{\rm c}(T)$, where $N_{\rm c}(T)$ is some 
%% critical value\footnote{At $T \gtrsim 0.5$, it was found
%% that $N_{\rm c}(T)\sim 6/T$ \cite{Anagnostopoulos:2007fw}.}
%% depending on $T$, 
%% the configuration thermalizes as the simulation proceeds
%% and we can calculate various observables by taking an average.
%% As $N$ approaches $N_{\rm c}(T)$, however, the equilibrium state becomes
%% meta-stable, and the elements of $X_i(t)$ start to increase
%% endlessly at some point in the simulation.

%%%%%%%%%%%%%%%%%%%%%%%%%%%%%%%%%%%%%%%%%%%%%%%%%%%%%%%%%%%%%%%%%%%%%%
%%%%%%%%%%%%%%%%%%%%%%%%%%%%%%%%%%%%%%%%%%%%%%%%%%%%%%%%%%%%%%%%%%%%%%
%%%%%%%%%%%%%%%%%%%%%%%%%%%%%%%%%%%%%%%%%%%%%%%%%%%%%%%%%%%%%%%%%%%%%%
\section{Numerical tests of the gauge/gravity duality}
\label{sec:numerical-tests}
%\hspace{0.51cm}
%%%%%%%%%%%%%%%%%%%%%%%%%%%%%%%%%%%%%%%%%%%%%%%%%%%%%%%%%%%%%%%%%%%%%%
%%%%%%%%%%%%%%%%%%%%%%%%%%%%%%%%%%%%%%%%%%%%%%%%%%%%%%%%%%%%%%%%%%%%%%
%%%%%%%%%%%%%%%%%%%%%%%%%%%%%%%%%%%%%%%%%%%%%%%%%%%%%%%%%%%%%%%%%%%%%%

In this section we provide numerical tests of
the gauge/gravity duality including finite $\lambda$ 
and finite $N$ corrections, which correspond to 
the $\alpha '$ and string loop corrections, respectively,
on the gravity side.
These corrections are discussed separately in 
sections \ref{sec:alpha-prime} and \ref{sec:finite_N}.

%\subsection{test at large $N$ --- $\alpha '$ corrections}
\subsection{Test including $\alpha '$ corrections}
\label{sec:alpha-prime}

In this section we provide a test of
the gauge/gravity duality in the large-$N$ limit.
For that purpose, we perform Monte Carlo simulation 
of the D0-brane quantum mechanics (\ref{YMaction})
at large $N$ and compare the results with
the prediction (\ref{eq:E_alpha}) obtained on the gravity side.
%Despite the instability at finite 
As long as $N > N_{\rm c}(T)$, where $N_{\rm c}(T)$ is some 
critical value depending on $T$, 
the instability mentioned in the previous section
does not show up practically during Monte Carlo simulation,
%the configuration thermalizes as the simulation proceeds
and we can calculate various observables by taking an average
in a straightforward manner.
The critical value is found to behave as
$N_{\rm c}(T)\sim 6/T$ 
at $T \gtrsim 0.5$ \cite{Anagnostopoulos:2007fw}, 
which makes the lower $T$ region difficult to study.

\begin{table}[tb]
\centering 
\begin{tabular}{|c||c|c|c|c|c|c|c|}
%\begin{eqnarray}
%\begin{array}{|c|c||c|c|c|c|c||c|}
\hline
$T$ & $\Lambda=2$ & $\Lambda=3$ & $\Lambda=4$ & 
$\Lambda=5$ & $\Lambda=6$ & $\Lambda=7$ & $\Lambda=8$ \\ 
\hline
1.0 & 3.489(33) & 3.350(37) & 3.217(41) & 3.212(38) & 3.172(35) & 3.184(41) & 3.138(53)\\
\hline
0.9 & 2.978(14) & 2.810(18) & 2.722(20) & 2.659(33) & 2.637(26) & 2.651(43) & 2.666(26)\\
\hline
0.8 & 2.498(13) & 2.316(16) & 2.222(16) & 2.146(24) & 2.114(24) & 2.133(43) & 2.132(25)\\
\hline
0.7 & 2.054(11) & 1.868(15) & 1.758(12) & 1.689(24) & 1.647(25) & 1.627(27) & 1.628(24) \\
\hline
0.6 & 1.675(11) & 1.450(12) & 1.342(17) & 1.284(24) & 1.208(23) & 1.171(31) & 1.209(21) \\
\hline
0.5 & 1.368(9) & 1.128(12) & 1.005(13) & 0.948(22) & 0.883(19) & 0.857(27) & 0.872(19)\\
\hline
%\end{array}
%\end{eqnarray}
\end{tabular}
\caption{
The results for the internal energy $E/N^2$ obtained with $N=16$ 
at each $T$ and $\Lambda$. 
%In the right most column, we also present
%the results in the continuum limit obtained by extrapolation to $\Lambda=\infty$.
%The expectation values of 
%$N=16$, $\Lambda=2,3,4,5,6,7,8$ and $T=0.5,0.6, 0.7, 0.8, 0.9, 1.0$. 
}
\label{table1}
\end{table}

\begin{table}[tb]
\centering 
\begin{tabular}{|c||c|c|}
%\begin{eqnarray}
%\begin{array}{|c|c||c|c|c|c|c||c|}
\hline
$T$ &$\Lambda=\infty$, const.\ fit &$\Lambda=\infty$, linear fit \\ 
\hline
\hline
$1.0$ & $3.169(24)$ & $3.091(82)$ \\
\hline
$0.9$ & $2.651(17)$ & $2.680(86)$\\
\hline
$0.8$ & $2.124(16)$ & $2.104(75)$\\
\hline
$0.7$ & $1.634(14)$ & $1.565(123)$\\
\hline
$0.6$ & $1.201(14)$ & $1.203(110)$\\
\hline
$0.5$ & $0.873(12)$ & $0.832(98)$ \\
\hline
%\end{array}
%\end{eqnarray}
\end{tabular}
\caption{
The results 
for the internal energy $E/N^2$
in the continuum limit obtained by extrapolation to $\Lambda=\infty$
with the constant fit and the linear fit.
The constant fit is performed using the data within $6 \le \Lambda \le 8$,
whereas the linear fit is performed by fitting the data to 
$E/N^2 = a/\Lambda+b$ 
with the fitting range $4 \le \Lambda \le 8$ for $T=1.0$, 
$5 \le \Lambda \le 8$ for $T=0.9$, $0.8$ and $6 \le \Lambda \le 8$ 
for $T=0.7$, $0.6$, $0.5$. 
}
\label{table2}
\end{table}

%\FIGURE{
\begin{figure}[tbp]
\centering % \begin{center}/\end{center} takes some additional
           % vertical space
%\includegraphics[width=55mm,angle=-90]{N16_constant_vs_linear_3.eps}
\includegraphics[width=80mm]{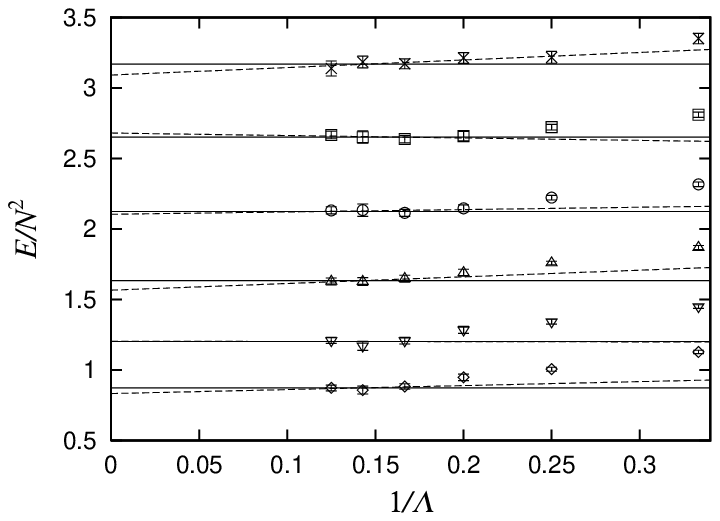}
\includegraphics[width=80mm]{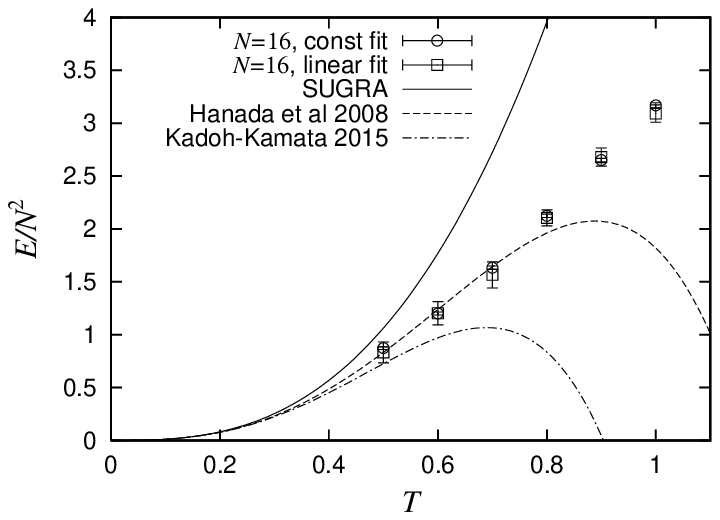}
%\includegraphics[width=55mm,angle=-90]{N16_constant_vs_quadratic.eps}
%
%\includegraphics[width=55mm,angle=-90]{N16_energy.eps}
%\includegraphics[width=55mm,angle=-90]{N16_deviation_from_SUGRA.eps}
%%
%%% See Hanada's mail from 10/17.%%%%%%%%%%%
%\includegraphics[width=5cm,angle=-90]{energy_N17_v1.eps}
%\includegraphics[width=5cm,angle=-90]{energy_N17_v2.eps}
    \caption{\label{fig:energy-N16-extrapol}
(Left) The internal energy $E/N^2$ for $N=16$ is plotted against $1/\Lambda$
for $T=0.5, 0.6, 0.7, 0.8, 0.9, 1.0$ from the bottom to the top.
The solid and dashed lines represent fits to $E/N^2 = {\rm const}.$
and $E/N^2 = a + b/\Lambda$, respectively.
%(Right) The lines represent fits to 
%$E/N^2 = a + b/\Lambda + c/\Lambda^2$
%and $E/N^2 = {\rm const}.$.
(Right)
The internal energy obtained by extrapolation to $\Lambda=\infty$.
The solid line represents 
the prediction of type IIA supergravity $E/N^2=7.41T^{2.8}$. 
The dashed line represents
a fit $E/N^2=7.41\,T^{2.8} + a T^{4.6}$
with $a=-5.58(1)$ obtained in ref.~\cite{Hanada:2008ez}.
%with the leading $\alpha '$ correction 
The dash-dotted line represents
a fit $E/N^2=7.41 \,T^{2.8} + a \,T^p$
with  $a=- 0.90(26) \times 10$ and $p=4.74(35)$
obtained in ref.~\cite{Kadoh:2015mka}.
%with the data obtained at lower $T$ with $N=32$ using a lattice formulation.
} 
\end{figure}
% %%%%%%%%%%%%%%%%%%%%%%%%%%%%%%%%%%%%%%%%%%%%%%%%%%%%%%%%%%%%%%%%%%%%%

In ref.~\cite{Hanada:2008ez},
the results obtained by Monte Carlo simulation
with $N \le 17$ and $\Lambda \le 8$ were compared
with the prediction including $\alpha '$ corrections.
%This analysis has been carried out first in ref.~\cite{Hanada:2008ez}
The numerical data were fitted by
an ansatz $E/N^2=7.41 \,T^{2.8} + a \,T^p$
%with $p=4.58 \pm 0.03$ and $a=-5.55 \pm 0.07$, 
with $p=4.58(3)$ and $a=-5.55(7)$, 
which is consistent with the prediction $p=4.6$
from the gravity side.
However, a recent paper 
\cite{Kadoh:2015mka} repeated the analysis including data points
at lower $T$ with $N=32$ using a lattice formulation.\footnote{The lattice size 
was $L=16$, which roughly corresponds to $\Lambda=8$ 
from the viewpoint of the number of degrees of freedom.}
The values obtained from the same fit was 
%$a=- 9.0\pm 2.6$ and $p=4.74\pm 0.35$. 
$a=- 0.90(26) \times 10$ and $p=4.74(35)$. 
%\cite{Kadoh:2012bg,Kadoh:2014hsa}.
In ref.~\cite{Hanada:2008ez},
a one-parameter fit with the power $p=4.6$ fixed was
also performed, and 
%the coefficient was determined as $a=-5.58 \pm 0.01$.
the coefficient was determined as $a=-5.58(1)$.

In order to clarify this discrepancy, 
we improve our previous analysis in ref.~\cite{Hanada:2008ez}
by making extrapolations to $\Lambda=\infty$.
In Fig.~\ref{fig:energy-N16-extrapol} (Left) we plot our results
obtained for $N=16$ against $1/\Lambda$.
%(We present the explicit values in Table \ref{table2}.)
We tried both a linear extrapolation
% (Left) 
$E/N^2 = a + b/\Lambda$
%and a quadratic extrapolation (Right) as well as 
and a constant fit.
The values at each $\Lambda$ and the values obtained
by the extrapolations are given
in Table \ref{table1} and \ref{table2}, respectively.
%In Fig.~\ref{fig:energy-N16} (Left),
In Fig.~\ref{fig:energy-N16-extrapol} (Right),
we plot the internal energy obtained by the extrapolations.
The new results are consistent with the fit
$E/N^2=7.41\,T^{2.8}-5.58\,T^{4.6}$
obtained in ref.~\cite{Hanada:2008ez}.

%\FIGURE{
\begin{figure}[tbp]
\centering % \begin{center}/\end{center} takes some additional
           % vertical space
\includegraphics[width=80mm]{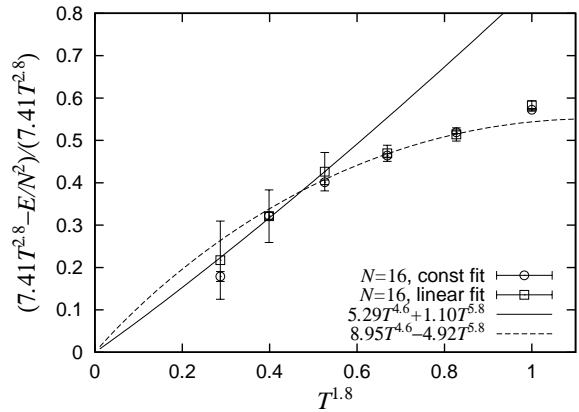}
%% \includegraphics[width=5cm,angle=-90]{N16_extrapolation.eps}
%% \includegraphics[width=5cm,angle=-90]{N16_energy.eps}
%% \includegraphics[width=5cm,angle=-90]{test.eps}
%%
%%% See Hanada's mail from 10/17.%%%%%%%%%%%
%\includegraphics[width=5cm,angle=-90]{energy_N17_v1.eps}
%\includegraphics[width=5cm,angle=-90]{energy_N17_v2.eps}
    \caption{\label{fig:energy-N16}
%(Right) 
The quantity $y=(7.41\,T^{2.8}-E/N^2)/7.41\,T^{2.8}$
representing the deviation of the internal energy from
the leading prediction from supergravity
is plotted against $x=T^{1.8}$.
The solid line and the dashed line
represent fits to $ y = a_1 x + a_2 x^{5/3}$
using the data 
within the range $0.5 \le T \le 0.7$ and $0.5 \le T \le 0.9$, 
respectively, obtained by the linear fit.
} 
\end{figure}
% %%%%%%%%%%%%%%%%%%%%%%%%%%%%%%%%%%%%%%%%%%%%%%%%%%%%%%%%%%%%%%%%%%%%%

In Fig.~\ref{fig:energy-N16}
%(Right) 
we plot
the difference between the obtained internal energy
and the leading prediction (\ref{eq:E_sugra}) from supergravity.
%without $\alpha '$ corrections. 
The difference is normalized by the leading prediction
as $y=(7.41\,T^{2.8}-E/N^2)/7.41\,T^{2.8}$ 
and it is plotted against $x=T^{1.8}$.
The leading $\alpha '$ corrections correspond to a linear behavior
towards the origin. 
Indeed we see a linear behavior for $T\le 0.7$
consistent with the fit obtained in ref.~\cite{Hanada:2008ez}.
On the other hand, the subleading terms are expected to show up as 
$y= a_1 x + a_2 x^{5/3} + a_3 x^{2} + a_4 x^{7/3} + \cdots$.
The solid line and the dashed line are 
fits to $y= a_1 x + a_2 x^{5/3}$ using the data points within 
the range $0.5 \le T \le 0.7$ and $0.5 \le T \le 0.9$, respectively,
obtained by the linear fit.
In the latter case, 
%the value of $a_1$ obtained by the fit 
%becomes much larger than the value 5.58 obtained
%in ref.~\cite{Hanada:2008ez}.
the left-most data point obtained by the constant fit
%, which corresponds to $T=0.5$,
is slightly off the fitting curve.
However, this may be due to finite-$N$ effects,
which become more significant at lower temperature
as is suggested from the $1/N$ expansion (\ref{eq:IIAene}).
%
%the data in a wider range of $T$
%
%$a_1=9.41$ and $a_2 = -5.41$
%if we neglect the data point at $T=0.5$.
%By fitting the results within the range 
%In fact, it is conceivable from the $1/N$ expansion (\ref{eq:IIAene})
%that finite $N$ effects become more significant at lower temperature.
In order to decide which fit is more appropriate, we clearly need
more data at lower temperature with larger $N$.

\subsection{Test including string loop corrections}
%\section{Numerical test of gauge/gravity duality at finite $N$} 
\label{sec:finite_N}
%%%%%%%%%%%%%%%%%%%%%%%%%%%%%%%%%%%%%%%%%%%%%%%%%%%%%%%%%%%%%%%%%%%%%%

In this section we test the gauge/gravity duality including
string loop corrections.
For that purpose, we need to study
the D0-brane quantum mechanics at small $N$ such as
$N=3,4,5$.
As is mentioned at the end of section \ref{sec:large-N},
the system has instability at small $N$ associated with
the flat directions in the action.
%However, the system becomes meta-stable at sufficiently low $T$.
In order to probe the instability, we define
%% \footnote{Instead of 
%% (\ref{R2-def}), we could have defined 
%% $R^2 =\frac{1}{N}
%%   \max_{0 \le t \le \beta} \sum_{i=1}^9{\rm Tr} \big(X_i(t)^2 \big)$
%% so that the constraint $R^2 \le R^2_\text{cut}$ we introduce below
%% may implement literally the condition ${\cal O} \le R^2_{\rm cut}$
%% %in the argument given 
%% discussed in section \ref{sec:large-N}. 
%% %We use (\ref{R2-def}) simply because it is well studied 
%% %in the literature \cite{Kawahara:2007fn}
%% We consider that this
%% does not make much difference because 
%% the fluctuation of $\sum_{i=1}^9{\rm Tr}\big(X_i(t)^2 \big)$ 
%% as a function of $t$ is typically small.}
\begin{align}
  R^2 \equiv \frac{1}{N\beta}
  \int_0 ^\beta  dt \sum_{i=1}^9{\rm Tr} \big(X_i(t)^2 \big) \ ,
\label{R2-def}
\end{align}
which represents the extent of the eigenvalue distribution of $X_i$'s.
In Monte Carlo simulation,
we prepare the initial configuration
of $X_i(t)$ with small $R^2$
by giving each element a small Gaussian random number.
%The observable $R^2 = \frac{1}{N}{\rm Tr} (X_i)^2$ 
At sufficiently low $T$,
we observe that $R^2$ stabilizes as the simulation proceeds,
and fluctuates around some value for a while and then starts to diverge.
%Thus, the system becomes meta-stable at sufficiently low $T$.
%
%Thus, the system becomes meta-stable, and we can
%consider the thermodynamics in a practical sense.
%
%In order to avoid this problem,
This behavior motivates us to consider the partition function
(\ref{eq:Z-cutoff}), where $R^2_{\rm max}$ is replaced
by $R^2$ for simplicity.\footnote{We consider that this
does not make much difference because 
the fluctuation of $\sum_{i=1}^9{\rm Tr}\big(X_i(t)^2 \big)$ 
as a function of $t$ is typically small.}
What we do in practice is to add the potential term
\begin{align}
V = 
\left\{
\begin{array}{ll}
c \; 
%\theta(R^2-R^2_\text{cut}) 
\left|R^2-R^2_\text{cut}\right| & \mbox{for $R^2 \ge R^2_\text{cut}$} \ , \\
0 & \mbox{for $R^2 < R^2_\text{cut}$} \ , \\
\end{array}
\right.
\label{def-potential}
\end{align} 
to the action,
where $c$ and $R_{\rm cut}^2$ are some parameters to be chosen appropriately.
%% \begin{align}
%% V = c \; \theta(R^2-R^2_\text{cut}) 
%% \left|R^2-R^2_\text{cut}\right| \ ,
%% \label{def-potential}
%% \end{align} 
%% where $c$ and $R_{\rm cut}^2$ are constants 
%% and $\theta (x)$ is the Heaviside step function.
%The parameter $R_{\rm cut}^2$ plays the role of the IR cutoff.

%\FIGURE{
\begin{figure}[t]
\centering % \begin{center}/\end{center} takes some additional
           % vertical space
\includegraphics[width=80mm]{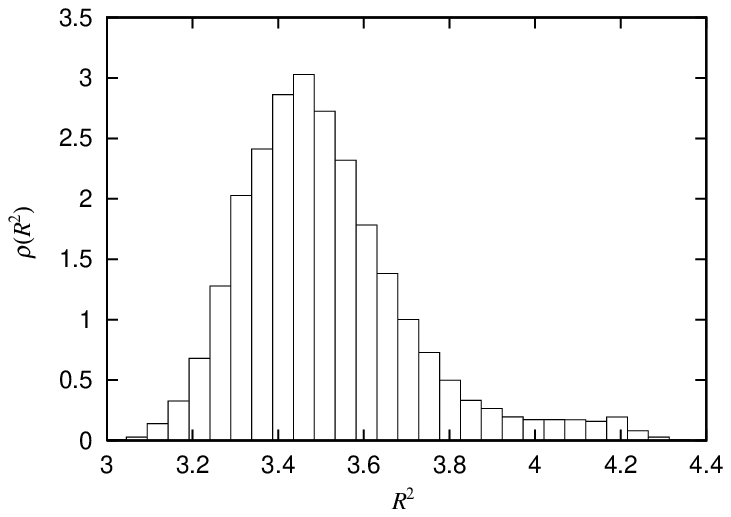}
\includegraphics[width=80mm]{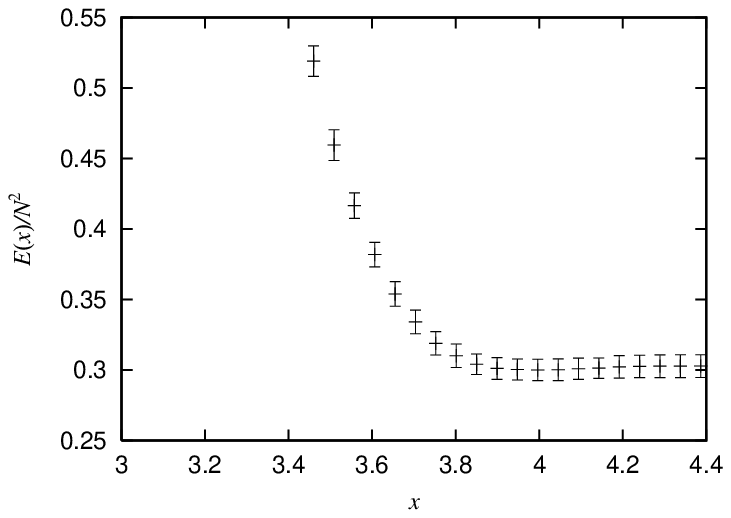}
\includegraphics[width=80mm]{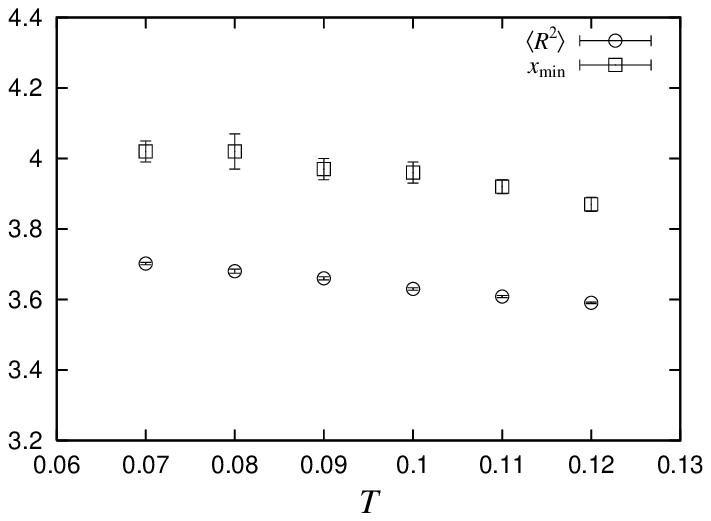}
    \caption{\label{histogram}
(Top-Left) The distribution of $R^2$ for $N=4$, $T=0.10$ and $\Lambda=10$
using $c=100$ and $R^2_\text{cut}=4.2$ in (\ref{def-potential}).
(Top-Right) The function $E(x)/N^2$ for the same set of parameters.
(Bottom) The extent of the bound states
estimated by $x_{\rm min}$ as described in the text is 
plotted as a function of $T$ for $N=4$ and $\Lambda=16$.
We also plot the expectation value $\langle R^2 \rangle$
obtained from configurations with $R^2 \le x_{\rm min}$.
} 
\end{figure}

Let us define the distribution of $R^2$ by
\begin{align}
\rho(x) = \Big\langle \delta(R^2-x) \Big\rangle \ ,
%% \rho(x) =
%% \left\langle \delta\left( \frac{1}{N\beta}
%%   \int_0 ^\beta  dt \sum_{i=1}^9{\rm Tr} \big(X_i(t)^2 \big) - x
%% \right) \right\rangle \ ,
\label{rho-def}
\end{align}
where the expectation value is taken in the system with 
the potential (\ref{def-potential}).
Figure \ref{histogram} (Top-Left) 
shows 
%the distribution of $R^2$
the distribution $\rho(x)$
obtained from Monte Carlo simulation with $N=4$, $T=0.10$, $\Lambda=10$, 
where we have set $c=100$ and $R^2_\text{cut}=4.2$.
%and the distribution function is normalized as $\int \rho(x)dx = 1$.
There is a clear peak around $R^2 \sim 3.5$,
which indicates the existence of the meta-stable bound states.
The long tail at $ R^2 \gtrsim 4$ represents the run-away behavior 
caused by the instability.
In Fig.~\ref{histogram} (Top-Right)
we plot the internal energy $E(x)/N^2$ obtained 
by averaging only over configurations with $R^2 < x$
for the same set of parameters.
We see a clear plateau around $x\sim 4$,
which confirms the argument given in section \ref{sec:large-N}.
%This motivates us to define the internal energy of the bound state 
%as the value of the local minimum of $E(x)/N^2$ in the plateau.
Practically, we define the internal energy of the bound states 
by the local minimum of $E(x)/N^2$ in the plateau region.
%% In Fig.~\ref{histogram} (Top-Right)
%% %Fig.~\ref{plateau}, 
%% we plot $E(x)/N^2$ for $N=4$, $T=0.10$, $\Lambda=10$, $c=100$ and 
%% $R^2_\text{cut}=4.2$.

The extent of the bound state can be identified
as the lower end of the plateau region,
which we denote as $x_{\rm min}$.
In practice, we obtain the value of $x$, at which
$E(x)/N^2$ deviates from the local minimum by 5\%,
and similarly the value of $x$ allowing 10\% deviation.
%In practice, we obtain the values of $x$, at which
%$E(x)/N^2$ deviates from the local minimum by 5\% and 10\%.
We use the average of the two values as an estimate of $x_{\rm min}$ and
the difference as an estimate of the ambiguity (``error'').
In Fig.~\ref{histogram} (Bottom)
we plot $x_{\rm min}$ thus obtained as a function of $T$
for $N=4$ 
and $\Lambda=16$
together with the expectation value $\langle R^2 \rangle$
obtained from configurations with $R^2 \le x_{\rm min}$.
We observe that both $x_{\rm min}$ and $\langle R^2 \rangle$
increase as $T$ is lowered.
Note that the quantity $\langle R^2 \rangle$
at large $N$ can be obtained without such a cutoff procedure,
and it is a monotonically increasing function of $T$.
(See Fig.~2 of ref.~\cite{Kawahara:2007ib}, for instance.)

%\FIGURE{
\begin{figure}[tbp]
\centering % \begin{center}/\end{center} takes some additional
           % vertical space
%\includegraphics[width=70mm]{c1.eps}
\includegraphics[width=80mm]{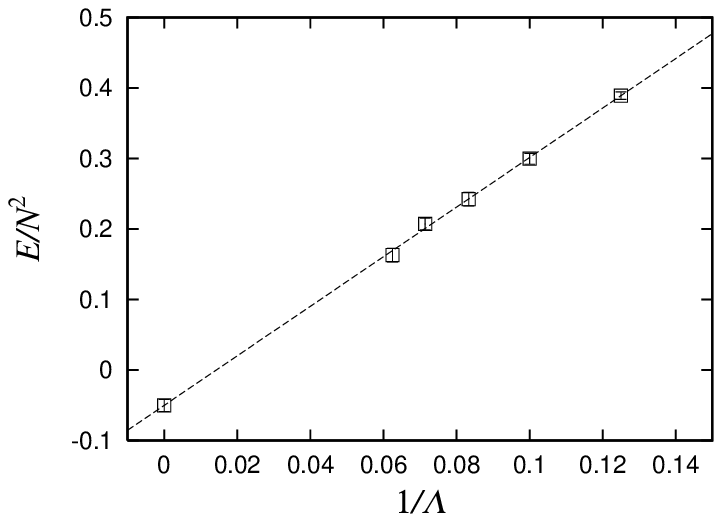}
\includegraphics[width=80mm]{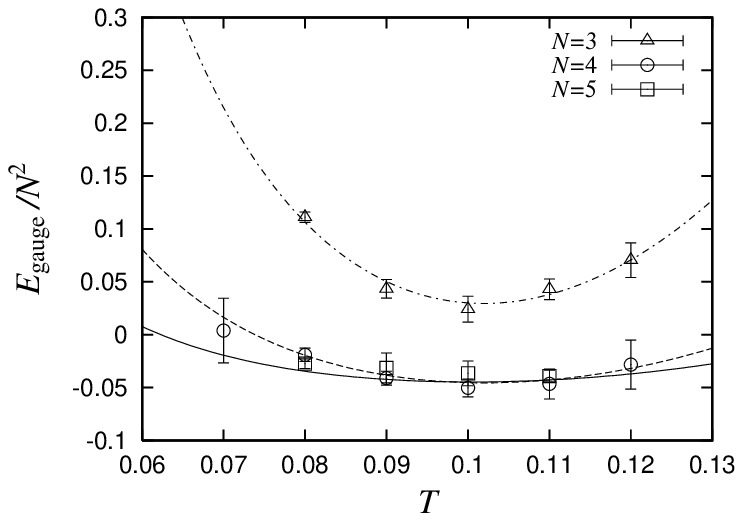}
    \caption{\label{alldata}
%% (Left) The parameter $c_1$ obtained by fitting our results to
%% $E_{\rm gauge}/N^2 = 7.41 T^{2.8} + c_1/N^2+c_2/N^4$
%% is plotted against $T$. The dotted line represents the behavior
%%   $f(T)=-5.76 \,T^{0.4}$ expected from the gravity side. 
%% (Right) 
(Left) The internal energy of the bound states
%defined as described in the text
is plotted against $1/\Lambda$ for $N=4$, $T=0.10$.
The straight line represents a fit to the behavior
$E=E_{\rm{gauge}}+ {\rm const.}/\Lambda$, 
and the value of $E_{\rm{gauge}}$ obtained by the fit
is plotted at $\Lambda=\infty$.
(Right)
The internal energy $E_\text{gauge}/N^2$ obtained from
our results by extrapolation to $\Lambda = \infty$
is plotted against $T$. The curves represent
$E_\text{gauge}/N^2= E_\text{gravity}/N^2 +(c \, T^{-2.6}+ \tilde{c} \, T^{p})/N^4$,
where the parameters $c$, $\tilde{c}$ and $p$ are obtained by fits
as described at the end of section \ref{sec:finite_N}.
The solid, dashed and dash-dotted lines correspond to
$N=5$, $4$ and $3$, respectively.
} 
\end{figure}
% %%%%%%%%%%%%%%%%%%%%%%%%%%%%%%%%%%%%%%%%%%%%%%%%%%%%%%%%%%%%%%%%%%%%%

Using the method explained above, 
we calculate the internal energy of the bound states for 
various $N$, $T$ and $\Lambda$. 
We have studied $0.08 \le T  \le 0.12$ for $N=3$, 
$0.07 \le T \le 0.12$ for $N=4$ and 
$0.08 \le T \le 0.11$ for $N=5$. 
%The cutoff parameter is chosen as $\Lambda =8,10,12,14,16$.
In Table~\ref{table}, 
we present our results for the internal energy obtained at 
each $N$, $T$ and $\Lambda$.
We make an extrapolation 
to $\Lambda = \infty$
% using results with $\Lambda =8,10,12,14,16$
assuming that finite
$\Lambda$ corrections to the internal energy
% start with a linear term in $1/\Lambda$ as 
are given by
$E=E_{\rm{gauge}}+ {\rm const.}/\Lambda$, 
from which we extract the value $E_{\rm{gauge}}$ 
in the continuum limit. 
This extrapolation is performed using
$ 8 \le \Lambda \le 16$ for $T\ge 0.10$ and 
$ 10 \le \Lambda \le 16$ for $T< 0.10$.
%$\Lambda=8,10,12,14,16$ for $T\ge 0.10$ and 
%$\Lambda=10,12,14,16$ for $T< 0.10$.
%Fig.~\ref{extrapo} 
Figure \ref{alldata} (Left) shows
% the extrapolation of $\Lambda$ for 
the case of $N=4$, $T=0.10$.
%, $c=100$ and $R^2_\text{cut}=4.2$.
In Fig.~\ref{alldata} (Right), 
we plot our results for $E_\text{gauge}/N^2$
obtained in the continuum limit by extrapolation to $\Lambda = \infty$.
(The explicit values 
%values of $E_{\rm{gauge}}/N^2$ 
are given
%presented 
in the right most column of Table~\ref{table}.)
The curves in this plot are explained at the end of this section.

%% \begin{figure}[tbp]
%%  \begin{minipage}{0.5\hsize}
%%   \begin{center}
%%    \includegraphics[width=70mm]{Nfourth.eps}
%%   \end{center}
%%   \caption{$(E_{\rm gauge} -E_{\rm gravity})/N^2$}
%%   \label{Nfourth}
%%  \end{minipage}
%%  \begin{minipage}{0.5\hsize}
%%   \begin{center}
%%    \includegraphics[width=70mm]{Egauge.eps}
%%   \end{center}
%%   \caption{$E_{\rm gauge}/N^2$}
%%   \label{Egauge}
%%  \end{minipage}
%% \end{figure}

\begin{table}[tb]
\centering 
\begin{tabular}{|c|c||c|c|c|c|c||c|}
%\begin{eqnarray}
%\begin{array}{|c|c||c|c|c|c|c||c|}
\hline
$N$ & $T$ & $\Lambda=8$ & $\Lambda=10$ & $\Lambda=12$ & 
$\Lambda=14$ & $\Lambda=16$ & $\Lambda=\infty$ \\ 
\hline
\hline
$5$ & $0.11$ & $0.353(9)$ & $0.273(13)$ & $0.225(17)$ & 
$0.181(19)$ & $0.160(29)$ & $-0.039(6)$ \\
\hline
$5$ & $0.10$ & $0.375(8)$ & $0.290(15)$ & $0.228(15)$ & 
$0.200(16)$ & $0.178(20)$ & $-0.037(11)$ \\
\hline
$5$ & $0.09$ & $0.397(6)$ & $0.323(8)$ & $0.269(10)$ & 
$0.218(10)$ & $0.193(13)$ & $-0.031(14)$ \\
\hline
$5$ & $0.08$ & $0.417(6)$ & $0.349(6)$ & $0.287(11)$ & 
$0.242(18)$ & $0.205(24)$  & $-0.027(5)$ \\
\hline
\hline
$4$ & $0.12$ & $0.366(6)$ & $0.297(7)$ & $0.242(11)$ & 
$0.213(11)$ & $0.153(10)$ & $-0.028(23)$ \\
\hline
$4$ & $0.11$ & $0.374(7)$ & $0.279(8)$ & $0.227(10)$ & 
$0.199(10)$ & $0.165(12)$ & $-0.047(14)$ \\
\hline
$4$ & $0.10$ & $0.389(5)$ & $0.300(8)$ & $0.242(10)$ & 
$0.207(8)$ & $0.163(10)$ & $-0.050(8)$ \\
\hline
$4$ & $0.09$ & $0.405(4)$ & $0.332(5)$ & $0.267(7)$ & 
$0.224(9)$ & $0.195(12)$ & $-0.041(7)$ \\
\hline
$4$ & $0.08$ & $0.422(4)$ & $0.365(5)$ & $0.298(7)$ & 
$0.254(9)$ & $0.223(11)$ & $-0.019(6)$\\
\hline
$4$ & $0.07$ & $0.442(3)$ & $0.375(4)$ & $0.329(4)$ & 
$0.289(5)$ & $0.245(6)$ & $0.004(30)$ \\
\hline
\hline
$3$ & $0.12$ & $0.407(6)$ & $0.327(8)$ & $0.295(9)$ & 
$0.264(13)$ & $0.243(14)$ & $0.071(16)$ \\
\hline
$3$ & $0.11$ & $0.397(6)$ & $0.332(8)$ & $0.284(8)$ & 
$0.238(11)$ & $0.220(10)$ & $0.043(10)$ \\
\hline
$3$ & $0.10$ & $0.396(4)$ & $0.323(7)$ & $0.280(7)$ & 
$0.242(10)$ & $0.201(8)$ & $0.024(12)$ \\
\hline
$3$ & $0.09$ & $0.411(5)$ & $0.344(5)$ & $0.293(8)$ & 
$0.257(6)$ & $0.240(17)$ & $0.049(8)$ \\
\hline
$3$ & $0.08$ & $0.426(4)$ & $0.355(9)$ & $0.313(7)$ & 
$0.288(15)$ & $0.263(14)$ & $0.111(5)$ \\
\hline
%\end{array}
%\end{eqnarray}
\end{tabular}
\caption{The results for the internal energy at each 
$N$, $T$ and $\Lambda$. In the right most column, we also present
the results in the continuum limit obtained by extrapolation to $\Lambda=\infty$.}
\label{table}
\end{table}

%\FIGURE{
\begin{figure}[tbp]
\centering % \begin{center}/\end{center} takes some additional
           % vertical space
\includegraphics[width=80mm]{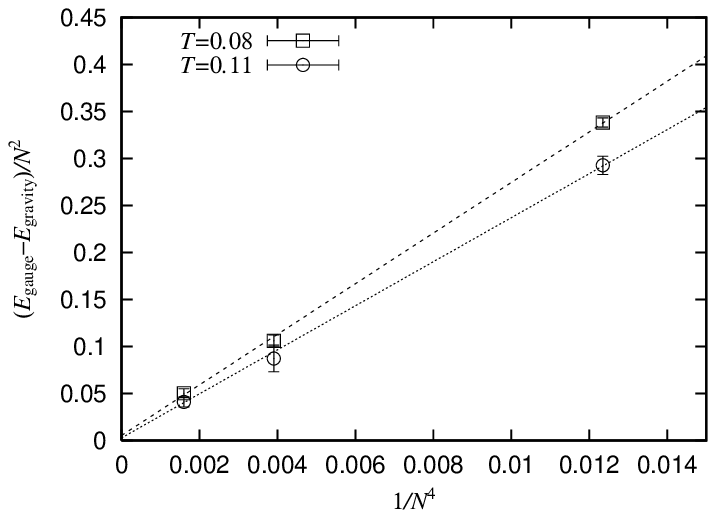}
\includegraphics[width=80mm]{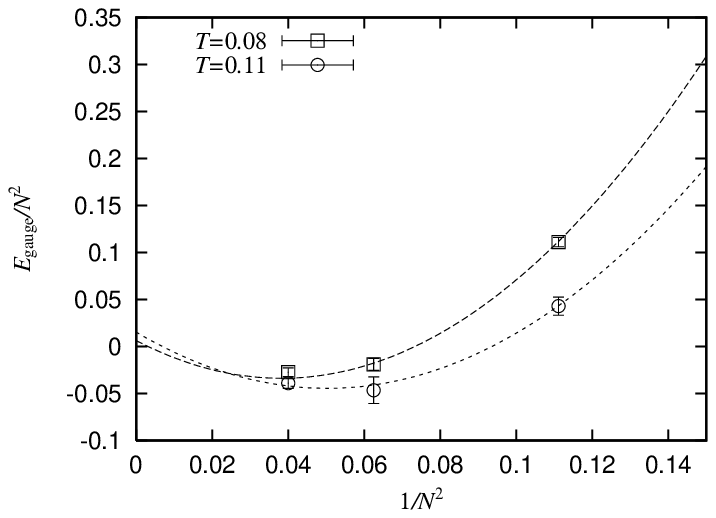}
    \caption{\label{Nfourth}
(Left) The difference $(E_{\rm gauge} -E_{\rm gravity})/N^2$
between the internal energy obtained by the gauge theory
and that predicted from the gravity side is plotted against $1/N^4$,
where the straight lines represent fits to a linear behavior.
(Right) The internal energy $E_{\rm gauge}/N^2$ obtained
by the gauge theory is plotted against $1/N^2$.
The dashed lines represent fits to the behavior
$E_\text{gauge}/N^2 =  7.41 \, T^{2.8} -5.76 \, T^{0.4}/ N^2
+{\rm const.}/N^4$, which is expected from the gravity side.
} 
\end{figure}

Let us compare our results with the prediction
%(\ref{finite N internal energy}) 
from the gravity side.
%When $0.08\le T\le 0.12$, 
Since the $\alpha '$ corrections 
%\cite{Hanada:2008ez}.
are negligible
in the temperature regime investigated here,
we can compare our results with (\ref{eq:finite_N}),
which we denote as $E_{\rm gravity}/N^2$ in what follows.
Let us also expand the internal energy 
of the bound states obtained on the gauge theory side as
\begin{align}
  \frac{E_{\rm gauge} }{N^2} 
= c_0(T)+\frac{c_1(T)}{N^2} + \frac{c_2(T)}{N^4} + \cdots \ .
\label{large N expansion}
\end{align}
If the gauge/gravity duality holds, 
the first and the second terms above
should coincide with those in 
(\ref{eq:finite_N}), namely, $c_0(T)=7.41 \, T^{14/5}$ and 
$c_1(T)=-5.77 \, T^{2/5}$.
In other words, the difference of the two quantities $E_{\rm gauge}$
and $E_{\rm gravity}$
%internal energies 
should behave as
\begin{align}
%\frac{E_{\rm gauge} }{N^2} - \frac{E_{\rm gravity} }{N^2} 
\frac{1}{N^2} ( E_{\rm gauge}  - E_{\rm gravity} )
=  \frac{c_2(T)}{N^{4}} + O\left(\frac{1}{N^6}\right) \ .
\label{difference}
\end{align}
In Fig.~\ref{Nfourth} (Left),
we plot $(E_{\rm gauge} -E_{\rm gravity})/N^2 $
against $1/N^4$, which can be nicely
fitted by a straight line passing through the origin.
We also observe a similar behavior for 
other values of $T$.
% with the same accuracy.
Thus, we confirm the behavior (\ref{difference}),
which implies that
%the equivalence of the internal energy
the gauge/gravity duality
holds including the leading quantum gravity correction.
In Fig.~\ref{Nfourth} (Right),
%Fig~\ref{Egauge}, 
we plot $E_\text{gauge}/N^2$ against $1/N^2$,
which can be fitted nicely by 
$E_\text{gauge}/N^2 =  7.41 \, T^{2.8} -5.76 \, T^{0.4}/ N^2
+{\rm const.}/N^4$ as expected from the gauge/gravity duality.
On the other hand, we also find that the $O(1/N^4)$ term is 
actually comparable to the $O(1/N^2)$ term. 
%Though it is beyond the scope of this paper,
This is related to the fact that the coefficient of the
$1/N^4$ term grows at low $T$ as we see below.
%The reason why the $O(1/N^6)$ term turned out to be small
%in this situation is unclear.

%\FIGURE{
\begin{figure}[tbp]
\centering % \begin{center}/\end{center} takes some additional
           % vertical space
\includegraphics[width=80mm]{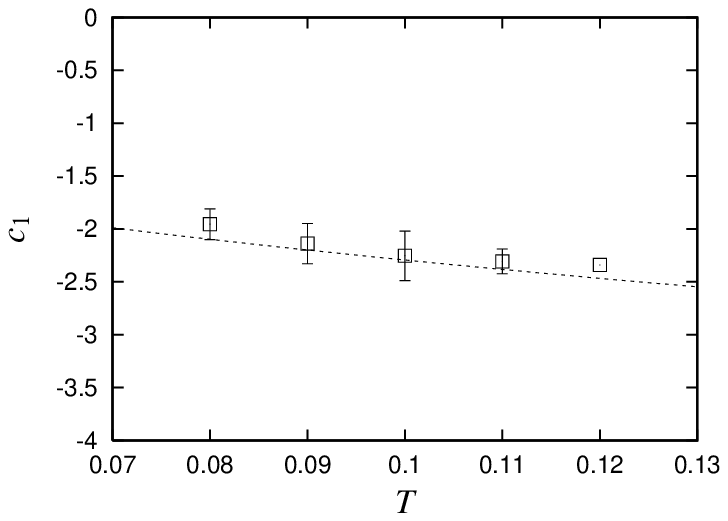}
\includegraphics[width=80mm]{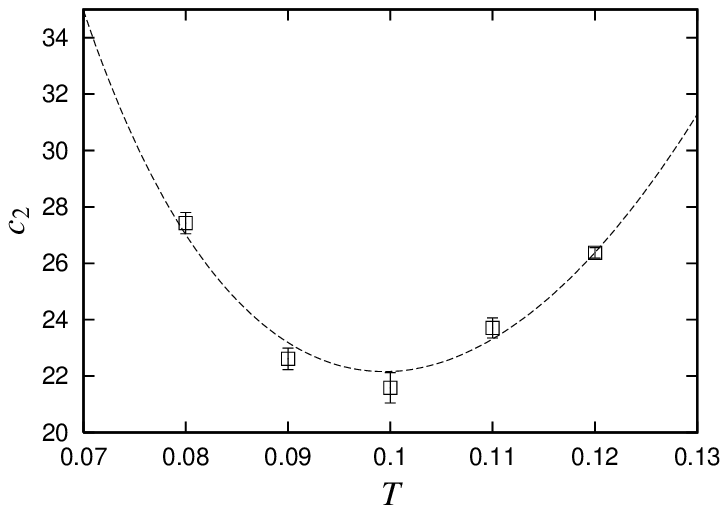}
    \caption{\label{c1}
(Left) The parameter $c_1$
obtained by fitting our results to
$E_{\rm gauge}/N^2 = 7.41 T^{2.8} + c_1/N^2+c_2/N^4$
is plotted against $T$. The dashed line represents the behavior
  $f(T)=-5.76 \,T^{0.4}$ expected from the gravity side. 
The data point for $T=0.12$ does not have an error bar since
we have only two data points ($N=3$ and $N=4$) for the two-parameter fit.
(Right) The parameter $c_2$ obtained 
by fitting our results to
$E_{\rm gauge}/N^2 = 7.41 T^{2.8} -5.76 \,T^{0.4}/N^2+c_2/N^4$
is plotted against $T$. The dashed line represents a fit
to the behavior $c_2 = c \, T^{-2.6}+ \tilde{c} \, T^{p}$ 
with $c= 0.032(2)$, $\tilde{c}=0.51(64)\times 10^5$ and $p=3.7(6)$,
%% (Right) The internal energy $E_\text{gauge}/N^2$ obtained from
%% our results by extrapolation to $\Lambda = \infty$
%% is plotted against $T$. The curves are obtained by fitting our results to
%% $E_\text{gauge}/N^2= E_\text{gravity}/N^2 +(c \, T^{-2.6}+ \tilde{c} \, T^{p})/N^4$.
} 
\end{figure}
% %%%%%%%%%%%%%%%%%%%%%%%%%%%%%%%%%%%%%%%%%%%%%%%%%%%%%%%%%%%%%%%%%%%%%

As an alternative analysis,
we fit our results for each $T$ with 
\begin{align}
E_{\rm gauge}/N^2 = 7.41 T^{2.8} + c_1/N^2+c_2/N^4 \ ,
\label{2-param-fit}
\end{align}
where $c_1$ and $c_2$ are fitting parameters.
%we can extract the coefficients $c_i(T)$ in the 
%$1/N^2$ expansion (\ref{large N expansion}) from our results at each $T$,
%and compare them with the predictions from the gravity side.
In Fig.~\ref{c1}, we plot the values of $c_1$ obtained by the two-parameter 
fit as a function of $T$, which agree well
with the behavior $c_1 = -5.76\, T^{0.4}$ expected from the gravity side.

Let us also discuss the $T$ dependence of $c_2$.
In Fig.~\ref{c1} (Right)
we plot the value of $c_2$ obtained
by fitting our data to (\ref{2-param-fit}) with $c_1 = -5.76\, T^{0.4}$ fixed.
From the prediction from the gravity side (\ref{general-formula}),
we find that $c_2 = c \, T^{-2.6}+\cdots$,
where the leading behavior is determined by the $(m,n)=(5,2)$ term 
in (\ref{general-formula}).
We can actually
fit our results
% for $c_2$ 
to
$c_2 = c \, T^{-2.6}+ \tilde{c} \, T^{p}$
%with $c=0.0340(11)$, $\tilde{c}=0.17(19)\times 10^6$ and $p=4.30(52)$,
with $c= 0.032(2)$, $\tilde{c}=0.51(64)\times 10^5$ and $p=3.7(6)$,
%where $c$, $\tilde{c}$ and $p$.
%% Therefore we consider that the $T$ dependence of $c_2$ is 
%% also consistent with the prediction from the gravity side.
where the second term $\tilde{c} \, T^{p}$ is meant to
represent all the subleading terms 
phenomenologically.\footnote{The value for $\tilde{c}$ 
obtained by the fit looks huge, but it is
actually compensated by the high power of $T$ within
the temperature region investigated here.}
Therefore we consider that the $T$ dependence of $c_2$ is 
also consistent with the prediction from the gravity side.
The curves in Fig.~\ref{alldata} (Right) represent
$E_\text{gauge}/N^2= E_\text{gravity}/N^2 +(c \, T^{-2.6}+ \tilde{c} \, T^{p})/N^4$ 
with $c$, $\tilde{c}$ and $p$ obtained above.
%We can see that the simulation data can be fitted very well by this function.

%%%%%%%%
%c               = 0.0323168        +/- 0.002079     (6.433%)
%\tilde{c}       = 50795.6          +/- 6.35e+004    (125%)
%p               = 3.73751          +/- 0.6014       (16.09%)
%%%%%%

%% \begin{figure}[htbp]
%%   \begin{center}
%%    \includegraphics[width=70mm]{c1.eps}
%%   \end{center}
%%   \caption{$c_1(T)$ is plotted against $T$. The curve represents 
%%   $f(T)=-5.76 \,T^{0.4}$.
%%   This plot appeared originally in ref.~\cite{Hanada:2013rga}. }
%%   \label{c1}
%% \end{figure}

%% \begin{figure}[htbp]
%%   \begin{center}
%%    \includegraphics[width=70mm]{alldata.eps}
%%   \end{center}
%%   \caption{$E_\text{gauge}/N^2$ is plotted against $T$. The curves are 
%% obtained by fitting the data points with $E_\text{gauge}/N^2= E_\text{gravity}/N^2 +(c \, T^{-2.6}+ \tilde{c} \, T^{p})/N^4$.
%% This plot appeared originally in ref.~\cite{Hanada:2013rga}. 
%% }
%%   \label{alldata}
%% \end{figure}

%%%%%%%%%%%%%%%%%%%%%%%%%%%%%%%%%%%%%%%%%%%%%%%%%%%%%%%%%%%%%%%%%%%%%%%%%%%%%%%%%%%%
%%%%%%%%%%%%%%%%%%%%%%%%%%%%%%%%%%%%%%%%%%%%%%%%%%%%%%%%%%%%%%%%%%%%%%%%%%%%%%%%%%%%
%%%%%%%%%%%%%%%%%%%%%%%%%%%%%%%%%%%%%%%%%%%%%%%%%%%%%%%%%%%%%%%%%%%%%%%%%%%%%%%%%%%%
\section{Summary and discussions} 
\label{sec:conclusion}
%%%%%%%%%%%%%%%%%%%%%%%%%%%%%%%%%%%%%%%%%%%%%%%%%%%%%%%%%%%%%%%%%%%%%%%%%%%%%%%%%%%%
%%%%%%%%%%%%%%%%%%%%%%%%%%%%%%%%%%%%%%%%%%%%%%%%%%%%%%%%%%%%%%%%%%%%%%%%%%%%%%%%%%%%
%%%%%%%%%%%%%%%%%%%%%%%%%%%%%%%%%%%%%%%%%%%%%%%%%%%%%%%%%%%%%%%%%%%%%%%%%%%%%%%%%%%%

In this paper we have performed numerical tests of
the gauge/gravity duality conjecture
including the $\alpha '$ and string loop corrections
by using 
$N$ D0-branes at finite temperature.
%the D0-brane system at finite temperature 
%as an explicit example.
On the gravity side, these are described by the black 0-brane geometry.
The leading part is obtained by the supergravity solution, and
the $\alpha '$ and string loop corrections
can be taken into account perturbatively,
which leads to 
the internal energy of the black 0-brane given by 
%eq.~(\ref{eq:IIAE}).
eq.~(\ref{eq:IIAene}).
On the gauge theory side,
the D0-branes are described by the matrix quantum mechanics,
which has been studied by Monte Carlo simulation.
%in order to investigate the strong coupling region 
%of the theory is relevant to the duality.

In the large-$N$ limit, 
the string loop corrections can be neglected
on the gravity side
and the internal energy is given by eq.~(\ref{eq:E_alpha}).
We have improved our previous analysis in ref.~\cite{Hanada:2008ez}
by performing extrapolations to the continuum limit.
While our new results are still consistent with the 
fit obtained previously, we have also suggested
an alternative fit obtained by taking into account
%the fit for the leading corrections may be 
the higher order $\alpha '$ corrections.
% may have to be taken into account.
%This reveals the importance of the continuum limit in the analysis, and 
More calculations at low temperature with larger $N$
%closer to the continuum limit 
are required for a definite conclusion to be reached.

We have also provided a test of 
the gauge/gravity duality including string loop corrections.
%This instability has its physical origins both on
%the gravity side and the gauge theory side.
In order to see the string loop corrections, we need to study
the D0-brane system with small $N$ such as $N=3,4,5$.
This is difficult because of the instability associated
with the flat directions of the potential.
At low temperature, however, we observe 
from Monte Carlo simulations that the bound states of D0-branes become meta-stable.
%meta-stable bound states of D0-branes, which 
We investigate the thermodynamics of these meta-stable bound states
by introducing a cutoff on the extent of the D0-branes.
%By measuring the internal energy as a function of the cutoff,
Indeed we find that the internal energy becomes independent of the cutoff
within a finite region.
% as expected from the meta-stability.
From this behavior, we were able to obtain the internal energy 
of the meta-stable bound states as a function of the temperature,
which turns out to be consistent with 
the result (\ref{eq:finite_N}) obtained recently 
on the gravity side including the leading string loop corrections.
To our knowledge, this is the first dynamical evidence that suggests
that the gauge/gravity duality holds at finite $N$.
(See ref.~\cite{Ardehali:2013gra} for studies on
kinematical aspects of the duality including $1/N$ corrections.)

A particularly interesting future direction would be to
investigate the D0-brane system at even lower temperature
%$T < T_{\rm c}\propto N^{-5/9}$ 
with larger $N$.
On the gravity side, it is expected that
the black 0-brane geometry becomes unstable,
% at larger $g_s$,
and undergoes a transition to a black hole moving along the eleventh 
direction \cite{Itzhaki:1998dd}
due to the Gregory-Laflamme instability \cite{Gregory:1993vy}.
%hyaku : added Gregory-Laflamme
This corresponds to a Schwarzschild black hole in eleven dimensions,
%which causes the Hawking radiation.
where the Hawking radiation becomes a non-negligible effect
unlike the case investigated in this paper.
The transition temperature has been obtained as
$T_{\rm c} = 0.574 N^{-5/9} + 0.707 N^{-11/9} + \cdots$
including the leading quantum correction \cite{Hyakutake:2015rqa}.
%hyaku : added 1 sentence and my paper
If one can see this transition in the dual gauge theory,
one should also be able to investigate the conjecture that
the same matrix quantum mechanics actually describes M theory 
nonperturbatively \cite{Banks:1996vh}.

%% Another important direction is to perform a precise comparison 
%% of the leading $\alpha'$ corrections.
%% On the gravity side, it is a challenging problem 
%% to derive the black 0-brane thermodynamics
%% including $\ell_s^6$ term. 
%% If this task is completed, the internal energy of the black 0-brane
%% is reliable in the wide range of $T$ and $N$. 
%% Then the numerical simulation is more tractable to
%% test the gauge/gravity duality conjecture for the D0-branes.

%%%%%%%%%%%%%%%%%%%%%%%%%%%%%%%%%%%%%%%%%%%%%%%%%%%%%%%%%%%%%%%%%%%%%%
%%%%%%%%%%%%%%%%%%%%%%%%%%%%%%%%%%%%%%%%%%%%%%%%%%%%%%%%%%%%%%%%%%%%%%
%%%%%%%%%%%%%%%%%%%%%%%%%%%%%%%%%%%%%%%%%%%%%%%%%%%%%%%%%%%%%%%%%%%%%%
\section*{Acknowledgements}
\hspace{0.51cm}
%%%%%%%%%%%%%%%%%%%%%%%%%%%%%%%%%%%%%%%%%%%%%%%%%%%%%%%%%%%%%%%%%%%%%%
%%%%%%%%%%%%%%%%%%%%%%%%%%%%%%%%%%%%%%%%%%%%%%%%%%%%%%%%%%%%%%%%%%%%%%
%%%%%%%%%%%%%%%%%%%%%%%%%%%%%%%%%%%%%%%%%%%%%%%%%%%%%%%%%%%%%%%%%%%%%% 
The authors would like to thank Sinya Aoki, Sean Hartnoll, 
Issaku Kanamori, Hikaru Kawai, Erich Poppitz, 
Andreas Sch\"{a}fer, Stephen Shenker,
Leonard Susskind, Masaki Tezuka, Akiko Ueda and Mithat \"{U}nsal for 
valuable discussions and comments.
M.~H.\ is supported by the Hakubi Center 
for Advanced Research, Kyoto University
and by the National Science Foundation 
under Grant No.\ PHYS-1066293.
%This work was supported in part by the National Science Foundation under 
%Grant No.~PHYS-1066293 and the hospitality of the Aspen Center for Physics.
M.~H.\ and Y.~H.\ are partially supported 
by the Ministry of Education, Science, 
Sports and Culture, Grant-in-Aid for Young Scientists (B), 
25800163, 2013 (M.~H.)  and 24740140, 2012 (Y.~H.).
The work of G.~I. is supported in part 
by Program to Disseminate Tenure Tracking System, MEXT, Japan.
The work of J.~N.\ is supported in part by Grant-in-Aid for Scientific Research
(No.\ 20540286, 23244057)
from Japan Society for the Promotion of Science.
Computations were carried out
on PC cluster systems in KEK and Osaka University Cybermedia Center 
(the latter being provided by 
the HPCI System Research Project, project ID:hp120162).

\appendix

%%%%%%%%%%%%%%%%%%%%%%%%%% Gravity Part %%%%%%%%%%%%%%%%%%%%%%%%%%%%%%

%%%%%%%%%%%%%%%%%%%%%%%%%%%%%%%%%%%%%%
%%%%%%%%%%%%%%%%%%%%%%%%%%%%%%%%%%%%%%
%%%%%%%%%%%%%%%%%%%%%%%%%%%%%%%%%%%%%%
\section{Hawking radiation in type IIA supergravity}
\label{appendix:hawking}

%%%%%%%%%%%%%%%%%%%%%%%%%%%%%%%%%%%%%%
%%%%%%%%%%%%%%%%%%%%%%%%%%%%%%%%%%%%%%
%%%%%%%%%%%%%%%%%%%%%%%%%%%%%%%%%%%%%%

The internal energy of the black 0-brane is 
affected by the Hawking radiation.
In order to estimate the strength of this effect, 
we consider the Stefan-Boltzmann law in 10-dimensional spacetime
\begin{alignat}{3}
  \frac{d\tilde{E}}{dt} = \sigma A_{\rm H} \tilde{T}^{10}
  = \sigma A_{\rm H} \lambda^\frac{10}{3} T^{10}\ , \label{eq:SB}
\end{alignat}
where $\sigma$ is some constant
and $A_{\rm H}$ is the area of the event horizon 
evaluated in the Einstein frame as
\begin{alignat}{3}
  A_{\rm H} &= 
e^{-2\phi} V_{S^8} \big( \ell_s H^\frac{1}{2} U \big)^8 \Big|_{\tilde{U}_0} 
  = \frac{2 (2\pi)^6\sqrt{\pi}}{7\sqrt{15}} \ell_s^{14} \lambda^2 U_0^\frac{9}{2}
  = \frac{2 (2\pi)^6\sqrt{\pi}}{7\sqrt{15}} 
a_1^{-\frac{9}{5}} \ell_s^{14} \lambda^2 T^\frac{9}{5} 
\end{alignat}
using the constant $a_1$ given in (\ref{eq:T_sugra}).
In terms of dimensionless quantities (\ref{eq:dim-less}), 
the energy loss per unit time is expressed as
% a function of temperature by
\begin{alignat}{3}
  \frac{dE}{d(\lambda^{1/3}t)} = 
\sigma' \big( \ell_s \lambda^\frac{1}{3} \big)^{14} T^{\frac{59}{5}} \ , \qquad
  \sigma' = \frac{2 (2\pi)^6\sqrt{\pi}}{7\sqrt{15}} a_1^{-\frac{9}{5}} \sigma \ .
\end{alignat}
Since the coefficient vanishes in the $\ell_s \to 0$ limit, 
the energy loss through the Hawking radiation can be neglected 
in the present calculation.
This conclusion is understandable since the near horizon limit 
corresponds to a particular case of the near extremal limit
as we mentioned below eq.~(\ref{near-horizon-limit}).

\section{Leading higher derivative corrections at one loop} %hyaku : \subsection -> \section
\label{appendix:one-loop}

In this appendix, we briefly review
the derivation \cite{H3} of the internal energy (\ref{eq:finite_N})
including the leading higher derivative correction 
at the one-loop level, which corresponds to the $(m,n)=(3,1)$ term
in the effective action (\ref{eq:IIAact}).

%% While the structure of the effective action (\ref{eq:IIAact}) is 
%% complicated and quite hard to handle.
%% There exists one exception, however, which is 
%% the leading higher derivative correction $(m,n)=(3,1)$ 
%% at the one-loop level. 

Including the correction, the 0-brane solution is modified and
given with an asymptotically flat metric as
\begin{alignat}{3}
  &ds^2 = - H_1^{-1} H_2^{\frac{1}{2}} F_1 dt^2 + H_2^{\frac{1}{2}} F_1^{-1} dr^2 
  + H_2^\frac{1}{2} r^2 d\Omega_8^2 \ , \quad
  e^\phi = H_2^\frac{3}{4} \ , \quad 
  C = \sqrt{1+\alpha^7} (H_2 H_3)^{-\frac{1}{2}} dt \ , \notag
  \\[-0.1cm]
  &H_i = 1 + \frac{r_-^7}{r^7} + \frac{\gamma}{r_-^6 \alpha^{13}} 
  \Big\{ h_i\Big(\frac{r}{r_-\alpha}\Big) 
+ \alpha^7 \hat{h}_i\Big(\frac{r}{r_-\alpha} \Big) \Big\} \ , \quad 
  F_1 = 1 - \frac{r_-^7 \alpha^7}{r^7} + 
\frac{\gamma}{r_-^6 \alpha^6} f_1\Big(\frac{r}{r_-\alpha}\Big) \ ,
\label{0-brane-loop-corrections}
\end{alignat}
where $\gamma = \frac{\pi^2}{2^{11} 3^2} \ell_s^6 g_s^2$.
The four functions $h_i(x)$ and $f_1(x)$ are uniquely 
determined as\footnote{In fact, the term $\frac{3747840}{x^7}$
in the function $f_1(x)$ was dropped in ref.~\cite{H2}
by imposing a stronger boundary condition,
although we actually need it to 
satisfy the asymptotic flatness of the solution.
%% In ref.~\cite{H2},
%% this extra term was dropped 
This does not affect
the final results for physical quantities such as entropy
%obtained in ref.~\cite{H2} are still correct
since they depend on $f_1(x)$ only through 
the combination $7 f_1(1) + f_1'(1)$.}
\begin{alignat}{3}
  h_1(x) &= \frac{1302501760}{9x^{34}}-\frac{57462496}{x^{27}}+\frac{12051648}{13
  x^{20}}-\frac{4782400}{13x^{13}}  \notag
  \\
  &\quad\,
  - \frac{3747840}{x^7} + \frac{4099200}{x^6} - \frac{1639680 (x-1)}{(x^7-1)} 
  + 117120 \Big( 18 - \frac{23}{x^7} \Big) I(x) \ , \notag
  \\[0.2cm]
  h_2(x) &= \frac{19160960}{x^{34}}-\frac{58528288}{x^{27}}+\frac{2213568}{13 x^{20}}
  -\frac{1229760}{13x^{13}} \notag
  \\
  &\quad\,
  - \frac{2108160}{x^7} + \frac{2459520}{x^6} 
  + 1054080 \Big( 2 - \frac{1}{x^7} \Big) I(x) \ , \notag
  \\[0.2cm]
  h_3(x) &= \frac{361110400}{9x^{34}} - \frac{59840032}{x^{27}}
  - \frac{24021312}{13x^{20}} - \frac{58072000}{13x^{13}} \label{eq:h}
  \\
  &\quad\, 
  - \frac{2108160}{x^7} + \frac{2459520}{x^6} 
  + 117120 \Big(18 - \frac{41}{x^7} \Big) I(x) \ , \notag
  \\[0.2cm]
  f_1(x) &= - \frac{1208170880}{9x^{34}} \!+\! \frac{161405664}{x^{27}} \!+\! \frac{5738880}{13x^{20}}
  \!+\! \frac{956480}{x^{13}} \!+\! \frac{3747840}{x^7} \!+\! \frac{819840}{x^7} 
I(x) \ , \notag
\end{alignat}
and the three functions $\hat{h}_i(x)$ are expressed as
\begin{alignat}{3}
  \hat{h}_1(x) &= \frac{1035722240}{9x^{27}} \!+\! \frac{1721664}{x^{20}} 
  \!+\! \frac{22955520}{13 x^{13}} \!+\! \frac{1912960}{x^6} 
  \!-\! 1639680 \frac{x-1}{x^7-1} \!+\! 234240 I(x) \ , \notag
  \\[0.2cm]
  \hat{h}_2(x) &= - \hat{h}_3(x) = 
  -\! \frac{94330880}{9 x^{27}} \!+\! \frac{655872}{x^{20}} 
  \!+\! \frac{13117440}{13 x^{13}} \!+\! \frac{2186240}{x^6}
  \!+\! 1873920 I(x) \ , \label{eq:hhat}
\end{alignat}
where
\begin{alignat}{3}
  I(x) &= \log \frac{x^7 (x-1)}{x^7-1} 
  - \sum_{n=1,3,5} \cos \frac{n\pi}{7} \log \Big( x^2 + 2 x \cos \frac{n\pi}{7} + 1 \Big) \notag
  \\[-0.1cm]
  &\quad\,
  - 2 \sum_{n=1,3,5} \sin \frac{n\pi}{7} \bigg\{
  \tan^{-1} \bigg( \frac{x + \cos \tfrac{n\pi}{7} }
  {\sin \tfrac{n\pi}{7}} \bigg) - \frac{\pi}{2} \bigg\} \ . \label{eq:I}
\end{alignat}

Using the black 0-brane solution including 
the leading quantum correction (\ref{0-brane-loop-corrections}),
we can obtain various quantities associated with the solution.
The Hawking temperature is given by
\begin{alignat}{3}
  \tilde{T} &= \frac{7 \alpha^\frac{5}{2}}{4\pi r_- \sqrt{1+\alpha^7}}
  \Big[ 1 + \frac{\gamma}{r_-^6 \alpha^6} 
  \Big\{ \Big( \frac{8}{7} - \frac{1}{2(1+\alpha^7)} \Big) f_1(\alpha)
  + \frac{1}{7} f_1'(1) - \frac{1}{2(1+\alpha^7)}h_1(1) \Big\} \Big]
\end{alignat}
up to the linear order in $\gamma$,
while the mass and the charge are evaluated, respectively, as
\begin{alignat}{3}
  \tilde{M} = \frac{V_{S^8}}{2\kappa_{10}^2} \big\{ r_-^7 (7 + 8 \alpha^7)
  - 16865280 \gamma \, r_- \alpha \big\} \ , \qquad
  \tilde{Q} = \frac{V_{S^8}}{2\kappa_{10}^2} 7 r_-^7 \sqrt{1 + \alpha^7} \ ,
\end{alignat}
which shows that the charge is not affected by quantum corrections.

After taking the near horizon limit, the dimensionless temperature becomes
\begin{alignat}{3}
  T &= a_1 U_0^\frac{5}{2}
  \big( 1 + \epsilon a_2 U_0^{-6} \big) \ , \quad
  a_2 = \frac{9}{14} f_1(1) + \frac{1}{7} f_1'(1) - \frac{1}{2}h_1(1) \ , \quad
  \epsilon = \frac{\pi^6}{2^7 3^2 N^2} \ ,
\end{alignat}
and the dimensionless internal energy becomes
\begin{alignat}{3}
  \frac{E}{N^2} &= \frac{V_{S^8}}{2\kappa_{10}^2 N^2 \lambda^\frac{1}{3}} 
  \bigg\{ \frac{1 + 8 \sqrt{1 + \alpha^7}}{1 + \sqrt{1 + \alpha^7}} r_-^7 \alpha^7
  - 16865280 \gamma \, r_- \alpha \bigg\} \notag
  \\
  &= \frac{18}{7^3} \, a_1^2 \, ( U_0^7 - 3747840 \,\epsilon \,U_0 ) \notag
  \\
  &= \frac{18}{7^3} \, a_1^2 \,
  \left\{ a_1^{-\frac{14}{5}} T^{\frac{14}{5}} 
  - \epsilon \left( \frac{14}{5} a_2 + 3747840 \right) 
     a_1^{-\frac{2}{5}} T^{\frac{2}{5}} \right\} \notag
  \\
  &= 7.41 T^\frac{14}{5} - \frac{5.77}{N^2} T^{\frac{2}{5}} \ . 
\label{finite N internal energy}
\end{alignat}
This is consistent with the result derived in ref.~\cite{H2}
using the first law of the black hole thermodynamics.

%%%%%%%%%%%%%%%%%%%%%%%

%==================================================================
\section{Derivation of the formula for the internal energy}
\label{sec:EO}
%==================================================================

In this appendix we derive the formula (\ref{energy-formula}),
which is used to calculate the internal energy by Monte Carlo simulation.
%The derivation in 
Let us first rewrite the internal energy 
$E=-\frac{\partial }{\partial\beta} \ln Z$ as
\beq
E = - \frac{1}{Z(\beta)}
\lim_{\Delta \beta \rightarrow 0 } 
\frac{Z(\beta ') - Z(\beta)}{\Delta \beta} \ ,
\label{E-cal}
\eeq
where $\beta ' = \beta + \Delta \beta$,
and represent $Z(\beta ')$ for later convenience as
\beq
Z(\beta ') = \int 
[{\cal D} A']_{\beta '}
[{\cal D} X ']_{\beta '}
[{\cal D} \psi ']_{\beta '}
 \, \ee^{- S'}  \ ,
\eeq
where $S'$ is obtained from $S$ given in (\ref{YMaction})
by replacing $\beta$, $t$, $A(t)$, $X_i(t)$, $\psi_\alpha(t)$
with $\beta '$, $t'$, $A'(t')$, $X_i '(t')$, $\psi_\alpha ' (t')$.
In order to relate $Z(\beta ')$ to $Z(\beta)$,
we consider the transformation
\beq
t ' = \frac{\beta '}{\beta} \, t \ ,
\quad
A '(t ') = \frac{\beta}{\beta '} \, A(t) \ ,
\quad
X_i ' (t ') = \sqrt{\frac{\beta '}{\beta}} \, X_i (t) \ ,
\quad
\ps'(t') = \ps(t) \ ,
\label{beta-transformation}
\eeq
where the constant factors are
motivated on dimensional grounds.
As for the path integral measure, we impose
%in particular
$ [ {\cal D} X ' ]_{\beta ' } = [ {\cal D} X]_{\beta}$,
$ [ {\cal D} \psi ' ]_{\beta ' } = [ {\cal D} \psi]_{\beta}$
and
$[ {\cal D} A ' ]_{\beta ' } = [ {\cal D} A]_{\beta} $,
which corresponds to subtracting the internal energy
for the free theory in the definition (\ref{E-cal}).
Under this transformation, 
the kinetic term in $S'$ reduces to that in $S$,
but the interaction term transforms non-trivially as
%% Note also that the kinetic term in the action
%% (\ref{action}) is invariant under this transformation.
%% The interaction term is not invariant, though,
%% and we obtain
\beqa
%--------------------------------------------------
\int_0^{\beta '} \!\!dt ' \, \tr
\Bigl( [X_i ' (t ' ),X_j ' (t ' )] \Bigr)^2 
&=& 
\left(\frac{\beta'}{\beta}\right)^3
\int_0^\beta  \!\!dt  \, \tr
\Bigl( [X_i  (t  ),X_j (t )] \Bigr)^2 \ ,\\
%--------------------------------------------------
\int_0^{\beta '}  \!\!dt  \, 
\tr 
\Bigl( \ps_\alpha (t ' ) 
% (\gm_i)_{\alpha\beta} 
[{X_i}' (t ' ) ,\ps'_\beta (t ' )] \Bigr)
&=& 
\left(\frac{\beta'}{\beta}\right)^{3/2}
\int_0^\beta  \!\!dt  \, 
\tr
\Bigl( \ps_\alpha 
%(\gm_i)_{\alpha\beta} 
[X_i(t),\ps_\beta(t)] \Bigr) \ .
%--------------------------------------------------
\eeqa
This gives us the relation
\beqa
\label{Z/Z}
Z(\beta')=Z(\beta) \left\{
1 -   \frac{\Delta \beta}{\beta}
\left(
%{\cal E}_{\rm b} + {\cal E}_{\rm f} 
3 \langle S_{\rm b, int} \rangle 
+ \frac{3}{2} \langle S_{\rm f, int} \rangle
%% \frac{3}{\beta} S_{\rm b, int}
%% + \frac{3}{2\beta} S_{\rm f, int}
\right)
+{\rm O}((\Delta \beta)^2)
\right\} \ ,
\eeqa 
where $S_{\rm b, int}$ and $S_{\rm f, int}$ represent
the bosonic and fermionic part of the interaction terms, respectively.
Plugging these into (\ref{E-cal}), we get
\beq
%-------------------------------
\label{def-E}
E
% - \frac{d}{d \beta} \log Z(\beta) 
= \frac{3}{\beta}
\left( \langle S_{\rm b, int} \rangle
+ \frac{1}{2} \langle S_{\rm f, int} \rangle \right)
%{\cal E}_{\rm b} +{\cal E}_{\rm f}
%
%\frac{3}{4}\, 
%\left(
%{\cal C}_1 +{\cal C}_2
%\right) 
\ .
\eeq
Thus we are able to express the internal energy $E$
in terms of the expectation values,
which can be calculated by Monte Carlo simulation. 
However, the second term $\langle S_{\rm f, int} \rangle$
is computationally demanding since it involves fermionic matrices.
This motivates us to rewrite it in terms of quantities involving 
bosonic matrices only.

For that purpose, we consider 
a change of variables $X_i(t) \mapsto e^{\epsilon} X_i(t)$
in the partition function.
The kinetic term $S_{\rm b, kin}$ 
for the bosonic matrices $X_i(t)$ in (\ref{YMaction}) transforms
as $S_{\rm b, kin} \mapsto e^{2 \epsilon} S_{\rm b, kin}$,
whereas the interaction terms transform
as $S_{\rm b, int} \mapsto e^{4 \epsilon} S_{\rm b, int}$
and $S_{\rm f, int} \mapsto e^{\epsilon} S_{\rm f, int}$.
The path integral measure for the bosonic matrices $X_i(t)$
transforms as 
%$dX \mapsto e^{9 \epsilon N^2 (2\Lambda+1)} dX$.
$dX \mapsto e^{9 \epsilon \{ N^2 (2\Lambda+1) - 1\} } dX$,
where the ``$-1$'' is due to the fact
that the trace part in the constant mode of $X_i(t)$
does not appear in the action and hence should be omitted.
Since the partition function should be invariant under the change of 
variables, we obtain the identity
\beq
9 \{ N^2 (2\Lambda +1) - 1 \} = 2 \langle S_{\rm b, kin}\rangle
+ 4 \langle S_{\rm b, int}\rangle + \langle S_{\rm f, int}\rangle \ .
\eeq
Solving this equation for $\langle S_{\rm f, int}\rangle$,
and plugging it into (\ref{def-E}), we obtain
the formula (\ref{energy-formula}).

When we add the potential (\ref{def-potential}) to the action,
the formula (\ref{energy-formula}) is modified by an extra term
coming from the potential.
However, in the parameter region investigated in this work,
the extra term turns out to be negligible as we briefly discuss below.
For this reason, we use the formula (\ref{energy-formula}) 
with or without the potential (\ref{def-potential}).

For simplicity, we consider introducing 
the constraint $R^2 \le R^2_\text{cut}$,
which corresponds to inserting an extra factor
$\theta (R^2_\text{cut}- R^2)$
in the integrand of the path integral (\ref{eq:Z}).
Under the transformation (\ref{beta-transformation}),
$R^2$ defined in (\ref{R2-def}) transforms as
\beq
\frac{1}{\beta '}
\int_0^{\beta '} \!\!dt ' \, \tr
 X_i ' (t ' )^2 
=
\left(\frac{\beta'}{\beta}\right)
\frac{1}{\beta }
\int_0^{\beta } \!\!dt  \, \tr
 X_i  (t  ) ^2  \ .
\eeq
Therefore, the extra factor transforms as
\beqa
%--------------------------------------------------
&~& \theta \! \left( R^2_{\rm cut} - 
\frac{1}{N\beta '}
\int_0^{\beta '} \!\!dt ' \, \tr
 X_i ' (t ' ) ^2 \right) 
\nonumber \\
&=& \!\!\!
\theta \! \left( R^2_{\rm cut} - 
\frac{1}{N\beta}
\int_0^{\beta} \!\!dt \, \tr
 X_i (t) ^2 \right)
- \frac{\Delta \beta} {\beta} R^2_{\rm cut} 
  \delta \! \left( R^2_{\rm cut} - 
\frac{1}{N\beta}
\int_0^{\beta} \!\!dt \, \tr
 X_i (t) ^2 \right) \ ,
\eeqa
omitting the higher order terms in $\Delta \beta$.
%+{\rm O}((\Delta \beta)^2)
Therefore, the relation (\ref{Z/Z}) is modified as
\beqa
\label{Z/Z-modified}
Z(\beta')=Z(\beta) \left\{
1 -   \frac{\Delta \beta}{\beta}
\left(
%{\cal E}_{\rm b} + {\cal E}_{\rm f} 
3 \langle S_{\rm b, int} \rangle 
+ \frac{3}{2} \langle S_{\rm f, int} \rangle
%% \frac{3}{\beta} S_{\rm b, int}
%% + \frac{3}{2\beta} S_{\rm f, int}
+  R^2_{\rm cut} \rho(R^2_{\rm cut})
\right)
+{\rm O}((\Delta \beta)^2)
\right\} \ ,
\eeqa 
where $\rho(x)$ is defined by (\ref{rho-def})
in the system with the constraint $R^2 \le R^2_\text{cut}$.
Thus, (\ref{def-E}) is modified as
\beq
%-------------------------------
\label{def-E-modified}
E
% - \frac{d}{d \beta} \log Z(\beta) 
= \frac{3}{\beta}
\left( \langle S_{\rm b, int} \rangle
+ \frac{1}{2} \langle S_{\rm f, int} \rangle \right)
+ \frac{1}{\beta} R^2_{\rm cut} \, \rho(R^2_{\rm cut}) \ .
\eeq

Let us evaluate the extra term in the case shown in
Fig.~\ref{histogram} (Top-Left).
Here we have
$N=4$, $\beta=1/T=10$, $R_{\rm cut}^2=4.2$ and 
$\rho(R_{\rm cut}^2)\sim0.2$,
and hence the quantity $E/N^2$ receives 
a contribution from the extra term of the order of
$\frac{4.2 \times 0.2}{10 \times 4^2} \sim 0.005$, 
which may be neglected in the scale of 
Fig.~\ref{alldata} (Left).
In general, the extra term is suppressed by
the factor $\rho(R^2_{\rm cut})$, which is small 
if the system is stable enough.
%as far as the meta-stability is good enough.

%%%%%%%%%%%%%%%%%%%%%%%%%%%%%%%%%%%%%%%%%%%%%%%%%%%%%%%%%
%%%%%%%%%%%%%%%%%%%%%%%%%%%%%%%%%%%%%%%%%%%%%%%%%%%%%%%%%

\end{document}